\documentclass[preprint,showkeys,preprintnumbers,amsmath,amssymb, longbibliography]{revtex4}

\usepackage{graphicx}
\usepackage{dcolumn}
\usepackage{bm}
\usepackage{setspace}
\usepackage{feynmf}
\usepackage{amsmath}
\usepackage{latexsym}
\usepackage{amsfonts}
\usepackage{amssymb}
\usepackage{color}

\newcommand{\Z}{\mathbb{Z}}

\newcommand{\C}{\mathbb{C}}
\newcommand{\R}{\mathbb{R}}
\newcommand{\F}{\mathbb{F}}

\begin{document}

\preprint{quant-ph}

\begin{center}
\Large{\textsc{Towards Topological Quantum Computation? -\\
Knotting and Fusing Flux Tubes}}

\end{center}

\author{Meagan B. Thompson}


\begin{abstract}
\small

Models for topological quantum computation are based on braiding and fusing anyons (quasiparticles of fractional statistics) in (2+1)-D.  The anyons that can exist in a physical theory are determined by the symmetry group of the Hamiltonian.  Specifically, any theory of anyons must have braiding and fusion rules that satisfy consistency conditions known as the pentagon and hexagon equations (collectively, the Seiberg-Moore Polynomial Equations.)  
Maclane's coherence theorem states these are in fact all that is required in order to achieve commutativity of all combinations of fusion and braiding (i.e. a consistent physical theory).  Two applications of the Hexagon Equation yield the Yang-Baxter Equation(YBE) familiar from statistical mechanics: $ \sigma_j \sigma_{j+1} \sigma_j = \sigma_{j+1} \sigma_j \sigma_{j+1} $ where the $\sigma_i$ are the abstract braid group generators.  It is an unsolved mathematical problem to determine in general all the matrix solutions to the YBE.  In the case that the Hamiltonian undergoes spontaneous symmetry breaking of the full symmetry group G to a finite residual gauge group H, however, solutions are given by representations of the quantum double $D(H)$ of the subgroup. The quasi-triangular Hopf Algebra $D(H)$ is obtained from Drinfeld's quantum double construction applied to the algebra $\textit{F}(H)$ of functions on the finite group H.  As a vector space, $D(H) = \textit{F}(H) \otimes \C[H] = C(H \times H)$ where $\C[H]$ is the group algebra over the complex numbers and $C(H \times H)$ is the space of $\C$-valued functions on $H \times H$.

     A major new contribution of this work is a program written in MAGMA to compute the particles (and their properties - including spin) that can exist in a system with an arbitrary finite residual gauge group, in addition to the braiding and fusion rules for those particles.  We compute explicitly the fusion rules for two non-abelian group doubles suggested for universal quantum computation: $S_3$ and $A_5$, and discover some interesting results, subsystems, and symmetries in the tables.  $SO(3)_4$ (the restriction of Chern-Simons theory $SU(2)_4$) and its mirror image are discovered as 3-particle subsystems in the 8-particle $S_3$ quantum double.  The tables demonstrate that both $S_3$ and $A_5$ anyons are all Majorana, but this is not the case for all finite groups.  In the appendices, the quantum doubles for the remaining nonabelian subgroups of $SO(3)$ - $S_4$, $A_4$, and $D_4$ (the second in the infinite family $D_n$) - are tabulated and analyzed.  In addition, 
     the probabilities of obtaining any given fusion product in quantum computation applications are determined 
     and programmed in MAGMA.  
     Throughout, connections to possible experiments are mentioned.

\end{abstract}

\keywords{topological quantum computation, quantum physics, condensed matter}

\maketitle

\section{Introduction}

\normalsize

	Richard Feynman in 1982 was the first to conceive of the idea of a computer based on quantum mechanics as potentially offering more computational power than computers based on just classical physics.~\cite{Feynman}
  As computer components are beginning to attain distances on which quantum mechanical effects can no longer be ignored, continued size reduction would necessitate the design of a system that thrives on quantum mechanical distance scales.  
In addition, subsequent developments in algorithms have shown that certain problems believed to be classically computationally intractable would be efficiently solvable on a quantum computer.  (Notably, quantum algorithms for factoring large numbers and finding discrete logs efficiently~\cite{shor} would be able to decipher RSA and Diffie-Hellman, a fact which would have radical implications for the security of data ranging from credit cards to electronic communication).  These algorithms would suggest even the asymptotic computational power of a device is inextricably linked to the underlying physical mechanisms governing its operation in this more general framework. (The Church-Turing thesis asserts that all classical devices lead to the same asymptotic complexity.)


The main idea of topological quantum computation is to exploit in an automatically scalable way the physics of the underlying system to obtain computations that are inherently protected from errors.  Thus the issue of quantum error correction on the software level is sidestepped to a certain extent as error-prevention is a physical property of the system (effectively hardwired).  
From the computational perspective, the Threshold Theorem (as discovered by A. Steane in~\cite{steane} and D. Aharonov and M. Ben-Or in~\cite{Or}) guarantees that errorless quantum computation can be achieved using quantum error correction as long as the original error rate is below a fixed threshold.  However, even if the current threshold estimate (1 part in $10^{-4}$) is attainable, the overhead required to accomplish the error correction procedure is impractical - in general, thousands of qubits are required in order to obtain the desired control over one.  As it stands, topological accuracy may even be necessary to achieve the threshold.

Information is stored in the global topology of the system and processed via fusion
and braiding.  Since the environment interacts locally and the global topology of the system is not effected by local perturbations, the system is inherently protected against decoherence.  


\section{(2+1)-dimensional Topological Gauge Theories Set the Stage for Quantum Computation}

The systems in which topology acts to our advantage are those that are effectively (2+1)-dimensional.  The striking accuracy that can be achieved in topological systems for even classical electronics applications is very promising for the potential accuracy of a topological quantum computer.  For example, the resistivity in the IQHE is accurate to 1 part in $10^9$~\cite{ady} - It was even selected by NIST as the standard of measurement for resistance.  Replicating that type of inherent accuracy in the actual computations of a quantum computer would be an immense advantage over standard classical computations, regardless of asymptotic computational complexity.

     In three spatial dimensions, particles can be classified as bosons or fermions according to whether they have integer or half-integer spin respectively - those are the only two possibilities.  However, F. Wilczek showed that in two dimensions the structure can be much richer and in fact quasiparticles
     (anyons) of more general spin and exchange statistics can exist.~\cite{wil}

    The basic mechanism behind Topological Quantum Computation is to interchange and fuse these anyons in a controlled manner to achieve computations.  The term any-ons, originally coined by F. Wilczek, refers to the fact that these quasiparticles are a generalization of bosons and
     fermions in which the particles in abelian theories acquire ``any" (rational) phase $\theta$ when they are interchanged.  Bosons correspond to $\theta=0$ and fermions correspond to $\theta=\pi$.  Stability restricts $\theta$ to be any $\textit{rational}$ value.  In non-abelian theories the fields can even transform according to a higher-dimensional representation.  It is in fact non-abelian theories hold the most promise for quantum computation.  Anyons have been observed in a few different contexts in nature, notably in observations of the Fractional Quantum Hall Effect (FQHE) and are postulated to appear in
     a number of other different models which have yet to be experimentally tested.

Any physical topological theory is specified by the braiding rules and fusion rules.  From the mathematical perspective, the braiding and fusion rules must satisfy certain consistency conditions known as the Seiberg-Moore Polynomial Equations (also known as the pentagon and hexagon equations).  Maclane's coherence theorem states these are in fact all that is required in order to achieve commutativity of all combinations of fusion and braiding (i.e. a consistent physical theory).  The two main operations required for topological quantum computation are these knotting and fusion operations.

\subsection{Knots}

Knot theory originally began with the suggestion of Lord Kelvin that the fundamental building blocks of matter were knots in the ether.  While the original incarnation of Kelvin's idea was not quite physically grounded, it spawned a research program in mathematics that led to many exciting discoveries.  As a result, knots have been extensively studied and tabulated.  The effort received a boost with computer algebra in the 1990s with, among others, J. Weeks's computer program to compute various knot invariants.~\cite{weeks}

With the advent of relativity, physicists showed that space and time are linked and thus spacetime in everyday experience is in fact inherently 4-dimensional.  However, a modification of Kelvin's idea may have an element of truth in that in (2+1)-dimensions a configuration of fundamental excitations/particles in particular systems can be described by a knot rather than the individual excitations/particles.  Specifically, knotting is one of the basic mechanisms by which fundamental excitations can be used to achieve topological quantum computation.  

Specializing to particular systems may on the surface appear less fundamental than considering the elementary particle constituents of matter in a vacuum.  However, it may have more to do with particle physics than it would seem on first inspection - In recent years, it has been shown that classical notions of the vacuum are false.  The vacuum is not a stationary medium characterized by the absence of particles and interaction, but rather it is permeated by the quark-antiquark QCD condensate and seething with virtual particle-antiparticle pair production.  In addition, it has been shown that the equations of electromagnetism can be obtained as an effective field theory, analogous to those in condensed matter.  Thus effective field theories may in fact be the only ultimate physical reality.  In any event, effective field theories play a key role in topological physics.



\subsection{Fusion}


Fusion rules are necessary to determine the experimental feasibility of any anyon proposal.  In particular, if different anyon configurations in the proposal have different cumulative long-range fusion rules, the environment will be able to distinguish the states, resulting in decoherence.  In addition, fusion is the primary means of readout in Kitaev's, Preskill's, and Mochon's suggestions for universal quantum computation based on finite groups.  Thus it is important to understand the fusion rules well in order to build anyon computers.  Furthermore, the anyon properties of spin, charge, and magnetic flux are determined by their particle types (superselection sectors) and these similarly are important to keep in mind in the design of experimental realizations of anyons.  To this end, this paper will culminate with an original program in MAGMA to determine the particles permitted by the theory, the properties of those particles, and the fusion rules given the arbitrary (finite) symmetry group of the Hamiltonian.  


In order to understand the program, it is necessary to recall the underlying physical mechanisms governing (2+1)-dimensional discrete gauge theories.  To this end, a motivational example of the Aharonov-Bohm Effect in quantum electrodynamics (QED) is briefly described as a transition to the mechanism behind how subgroups of the (truncated) braid group $B(n,m)$ ($n,m \in \Z_{\geq 1}$) of anyons give rise to particles with different charges and fluxes in 2D.~\cite{prop}  Bais and deWild showed that the appropriate mathematical language for discussing braid groups is quasi-triangular Hopf algebras in the quasitensor category structure.~\cite{prop}  The particles are determined by the irreducible representations of a particular quasi-triangular Hopf algebra obtained by Drinfeld's Quantum Double Construction.


In Section C, 
the abelian theories $\Z/N\Z $ and the non-abelian theories $S_3$ and $A_5$ will be analyzed in depth.  $S_3$ and $A_5$ are non-abelian finite subgroups of $SO(3) $ (and $SU(3)$) that could be key from the perspective of universal quantum computation.  The particles (defined by charges and magnetic fluxes) and the fusion rules (from the comultiplication of irreducible representations of the quantum double) that can exist in these theories will be determined explicitly using the MAGMA program provided in Appendix A.  In addition, the spins of the particles will be computed.  A number of interesting results are found from the resulting tables.  In particular, some subsystems are discovered that are interesting from either a mathematical or quantum computation perspective.  The emphasis is on theories that are physically relevant.  An overview of the MAGMA program and a summary of the main results is provided.  The actual MAGMA code is included in Appendix A and the tabulation of the quantum doubles for the next dihedral subgroup and the remaining non-dihedral subgroups of $SO(3) $ is completed in Appendix B. 

Finally, possible experimental implementations and applications to universal quantum computation will be discussed.  The probabilities of obtaining certain fusion products in the quantum computation scheme will be derived.  The resulting probabilities are implemented in the MAGMA code included in Appendix A.


%


\section{Motivation: Topological Effects in Physics}

Conservation laws are critical to many physics analyses as they often simplify calculations
dramatically and aid intuition.  No$\ddot{e}$ther's Theorem states that for every
symmetry of the Hamiltonian, there is a conserved current and provides an explicit
formula for calculating the conserved current.  However, it turns out that this list is not exhaustive -
In fact, topological conservation laws can arise that cannot be traced to an underlying
No$\ddot{e}$ther current.

There are two major manifestations of topological effects - in coordinate space, the Aharonov-Bohm
Effect, and in momentum space, the Berry phase.

\subsection{Aharonov-Bohm Effect}

As motivation, recall the Aharonov-Bohm effect in quantum electrodynamics.~\cite{aharonovbohm}
Initially, gauge fields were introduced into classical physics as auxiliary fields to aid calculations -
While they were present in the canonical formalism, they were not necessary and did not appear in the
classical equations of motion (Maxwell's equations in the case of electromagnetism).  However,
quantum mechanics promotes observables to operators and mandates the use of the canonical
formalism.  As a result, the topology of the gauge fields can indeed have observable physical consequences.  In particular, in quantum mechanics, electrons become waves which exhibit an interference pattern.  Aharonov and Bohm predicted that a change in the gauge fields (by for example, the introduction of a solenoid in between the two beams) will cause a shift in the interference pattern observed.  As quantum mechanics only permits local interactions among fields and the two electron beams do not pass through the region in which the gauge fields are nontrivial, the use of the vector potential in this
case is inescapable.

While first introduced as a Gedanken-experiment for demonstrating
the influence of topological defects on physical systems, the Aharonov-Bohm Effect has now been verified by many actual
physical experiments (including a 1998 experiment using magnetostatic metal rings~\cite{metalring}).  The typical experiment demonstrating
the Aharonov-Bohm effect is the 2 slit electron diffraction experiment in which electrons are fired at the slits one at a
time and the diffraction pattern on a screen a fixed distance away is measured.  A change in the interference pattern is
observed when a solenoid (ideally, infinitely long so the magnetic field completely vanishes outside the solenoid) is
placed in between the two slits.  Thus it is found that the vector potential itself has intrinsic physical meaning in this
effectively two dimensional setting, not just the curl as originally supposed in classical electrodynamics.

Formally, Feynman's path integral method can be applied to to the system at hand.  The phase for the interference term is determined by the path difference between the two electron beams.  The degree of the effect is thus determined by the integral of the vector potential around a loop surrounding the solenoid.

Given a non-simply connected system, if the gauge potential is changed by an amount $\delta A$, then the resulting change in the wave equation is only invariant if
$ \oint \delta A(r) \cdot d r = n \frac{h}{e_j} $ where $n$ is an integer.  Thus the fractional part of $\frac{e}{h} \oint A(r) \cdot d r $ has measurable physical consequences.

In broken gauge theories, the flux is quantized - In order to ensure the wave function is single valued,
the vector potential can only encircle the solenoid an integer number of times.  Note that this
is a purely topological effect as it is determined solely by the winding number
of the path.  (In mathematical terms: the path integral exponent corresponds to an element of the fundamental group of the
space .  Specifically, the solenoid is a puncture in the plane (the third dimension is trivialized) so $\pi_1(\R^2 \setminus 0 \times \R) = \pi_1(S^1) = \Z$.)
 Thus topology can play an influential role in determining quantum physics phenomena.  In fact, since topological
 interactions are by nature robust against local perturbations, harnessing this resource for quantum computation could
 lead to a fault-tolerant quantum computer.

\subsection{Flux Metamorphosis}

In this paper, the topological effects arising from Hamiltonians in (2+1)-dimensions with finite
symmetry groups will be considered. Let H be a nonabelian group.  The basic idea is as follows: Consider two particles with fluxes in the residual gauge group:
$h_1, h_2 \in H$.  Fluxes introduce a nontrivial holonomy on the set of ground states that effects any test particle whose path winds around the flux.  WLOG take all the nontrivial parallel transport to occur in a single strand,
the Dirac string attached to the particle.  (This corresponds to fixing the gauge.) The Dirac string will be
represented as a single thick line in the vertical direction for all particles.  %

Now, the long range flux of the two particles combined is $h_1 h_2$.  Exchanging the two particles counterclockwise, the
particle with flux $h_2$ passes through the Dirac string of the particle with flux $h_1$.  Thus the flux of the particle with
flux $h_2$ will change to a new flux, call it $h_2'$.  Interchanging the particles should not have an effect on their cumulative
long-range topological quantum numbers.  Thus, when the two particles are exchanged counterclockwise, the cumulative flux of
the two particles should remain equal to $h_1 h_2$.  Thus the relation $h_2' h_1 = h_1 h_2$ is obtained, or $h_2' = h_1 h_2 h_1^{-1}$.  This effect
is called flux metamorphosis and applies to any 2D topological theory based on a non-Abelian group.

\begin{figure}[t]
\centerline{\includegraphics[width=11cm, height=6cm]{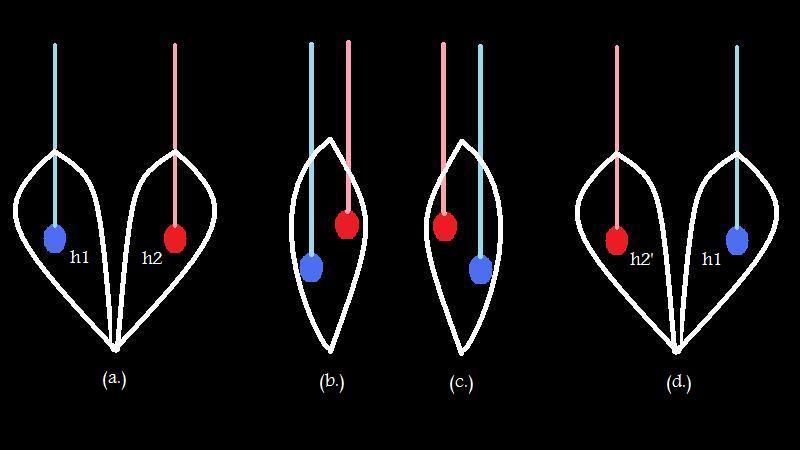}}
\caption[Flux Metamorphosis]{Two particles undergo flux metamorphosis when exchanged.  All nontrivial gauge transformations occur in the ``Dirac strings" of the particles, denoted here by the lighter lines going off to infinity. (a.) $h_1, h_2 \in H$ (b.)-(c.) $h_2$ is moved counterclockwise around $h_1$. (d) $h_2' = h_1 h_2 h_1^{-1}$.} %
\label{fig:fluxmeta}
\end{figure}

\section{Physics: Hamiltonian Description}

Modern physics has evolved closer to group theory as Hamiltonians can be classified solely on the basis of their symmetry
groups and particle content.  In fact, the current standard model is based on the Hamiltonian described by the gauge symmetry group
$SU(3)\times SU(2) \times U(1)$.  
The symmetry groups seen in nature are given by spontaneous symmetry breaking (SSB) of
this gauge group to subgroups.  $SU(3)$ is the strong force and $SU(2) \times U(1)$ is the electroweak force.  Electromagnetism
arises in SSB of the $SU(2) \times U(1)$ part to a $U(1)$ subgroup which involves both $SU(2)$ and $U(1)$
components.  (The $W^{\pm}$ bosons of the weak force are charged, so the $U(1)$ cannot be
cleanly separated from the $SU(2)$ to be interpreted as charge.)  In fact, superconductivity (also likely to be critical
for quantum computation) corresponds to the further SSB of the $U(1)$ electromagnetic gauge group to a subgroup.
This section will be devoted to understanding the underlying Physical foundations behind the existence of the particles.
First, SSB will be defined and a generic Hamiltonian for a physical system is provided. Then the restriction is made to the case of
$n$ particles in $(2+1)$-dimensional spacetime and the symmetry groups that can occur for this restricted
class of systems are determined.

\subsection{Hamiltonian}

Since the Hamiltonian of the system as a whole has to transform as a singlet under the representations of the symmetry group,
the representations limit the types of terms that can occur in the Hamiltonian.  In fact, the specification of a symmetry group
and the particles that are actually present determines the Hamiltonian uniquely.  In (2+1)-dimensional gauge theories based on
finite groups, the particles must satisfy consistency conditions so that braiding and fusion between three particles commute.
The consistency requirements restrict the types of particles that can possibly exist in the theory.

    To be explicit, the generic action in the Yang-Mills-Higgs theory can be written as
    $$ S = S_{YMH}+S_{matter}$$

In quantum field theory there are two types of symmetries:
\begin{enumerate}
\item global symmetries (physical symmetries of the theory, which have an associated conserved No$\ddot{e}$ther current) and
\item local gauge symmetries (nonphysical redundancies in the coordinates chosen, so do not have a conserved No$\ddot{e}$ther current).  \end{enumerate}

Suppose the action S (equivalently Hamiltonian, Lagrangian) has symmetry group G.  In the former case 1, if the global
symmetry is broken to a finite subgroup then according to Goldstone's theorem a massless particle (a Goldstone boson) is
created for every broken symmetry generator.  The emphasis in this thesis is on the latter case 2, in which the vector gauge fields (usually
originally massless) acquire mass through the Higgs mechanism.

\subsection{Symmetry Breaking} Starting with a Hamiltonian with gauge group G, there are two basic ways to break symmetry:

\begin{enumerate}
 \item Add an external potential: If the potential does not respect the full symmetry of the original Lagrangian, then
 the symmetry of the resulting Lagrangian becomes some subgroup of the original symmetry group.
 \item Spontaneous Symmetry Breaking (SSB): (described in most Quantum Field Theory textbooks, see for instance~\cite{wein}) - In this case, the full symmetry of the original Lagrangian
  is broken by the choice of a particular ground state (frequently this is referred to as the vacuum expectation value,
  v.e.v.) The order parameter space is isomorphic to G/H (an element of G/H rotates from one ground state to another in the
  system.)
\end{enumerate}

In the Standard Model of particle physics, the gauge boson acquires mass through the Higgs mechanism.  The same phenomenon occurs in Condensed Matter Physics in superconductivity - in fact, an (equivalent) formulation of superconductivity is that the photon field of electromagnetism acquires mass within the superconductor.

\subsection{Monopoles and Instantons}

Monopoles can arise as either Dirac or Polyakov-'t Hooft monopoles in field theories.  Both carry quantized flux, yet there are some crucial differences to keep in mind. (~\cite{cole} is a good reference.)  

\subsubsection{Dirac Monopoles}

P. Dirac predicted the existence of monopoles in the quantum theory of electromagnetism as a method of accounting for charge
quantization.  These Dirac monopoles are singular solutions of the field equations carrying
infinite energy.  In addition, Dirac monopoles have the aesthetic appeal of making Maxwell's equations symmetric.
To date, however, magnetic monopoles have not been observed in nature (Every bar magnet appears to split into
smaller bar magnets when torn apart down to the smallest scales measured).
  Dirac monopoles require the introduction of non-electromagnetic degrees of freedom,
  charged particles whose mass and spin are free parameters.

Mathematically, the monopoles are labeled by elements of the fundamental group ($\pi_1(G/H)$).

\subsubsection{Polyakov-'t Hooft Monopoles}

In broken gauge theories, it is not Dirac monopoles but rather Polyakov-'t Hooft monopoles that play a key role.  These are of a very
different nature than Dirac monopoles - they are non-singular solutions of the field equations that carry finite energy,
and necessitate no new degrees of freedom, i.e. all their parameters can be expressed in terms of the ordinary scalar and vector meson parameters.

Recall that $\pi_i(M), i \geq 2$ is always abelian.

\subsubsection{Instantons}

When the gauge group G is not simply connected, the lift $\bar{H}$ of the subgroup $H$ to the universal cover determines the different magnetic vortices.

\subsection{Higgs Screening}

In light of the preceding discussion of the Aharonov-Bohm effect, it is important to note that the physical charge de facto plays
two distinct roles in the Yang Mills Higgs theory. - It serves as both the coupling constant for the Coulomb interactions and as the
coupling constant for the Aharonov-Bohm scattering.
In the Higgs phase, Coulomb charges are screened at large distances (compared to the scale $\frac{1}{M_A}$, where $M_A$ is the
mass of the Higgs gauge fields).  However, topological interactions may persist at those distance scales.~\cite{higgscreen}
In order for a given charge $q$ to be detected via Aharonov-Bohm scattering, the screening charge $q'$ must not couple to the
Aharonov-Bohm interaction.

The resulting massive vector bosons mediate short-range
interactions only (the field strength decays exponentially as the mass of the Higgs field).  Since we are interested in the
global properties of the system, consider the long-distance (low energy) regime in which the Higgs field is condensed in one
of its ground states.  Since the field strength decays exponentially, the interactions between two particles due to their
fundamental gauge quantum numbers (throughout this paper the term electric charge will be used, by analogy with the electromagnetic case) at large
separation are exponentially suppressed.  (In the case of the electromagnetic field, this can be interpreted as Coulomb
screening by the Higgs medium.)

However, long range topological quantum numbers (which will henceforth be called flux) are still
well-defined and the system can thus be manipulated by the braiding and fusion determined by these topological quantum
numbers.  Since the resulting operations depend only on the topological parameters of the particles, the specific process or
path by which the braiding and fusion is carried out does not effect the conclusions.  (The framework described can thus be
applied to many different possible topological quantum computation implementations.  Since the computation is robust against
local deviations, this would also lead to reliable quantum gates.)

\subsection{Classification Based on the Symmetry Group}

\subsubsection{Three Dimensions}
Consider a physical system consisting of $n$ identical point particles.  (Note that for these particles to be truly
indistinguishable, the system has to be quantum mechanical.)  In three or more dimensions, any loop that circles twice
around a point can be continuously deformed to the identity  ($\pi_1(SO(3)) = \Z/2\Z$).  Thus the only physical property
corresponds to the particular permutation of the n particles (the winding number is always trivial), and the fundamental
group of the configuration space of the underlying manifold of this physical system is given by
$\pi_1(C_n(M)) = \pi_1((M^n-D)/S_n) \simeq S_n$, where D denotes the configurations in which two or more particles coincide.
Thus the Hamiltonian transforms under a representation of the symmetric group.  The symmetric group $S_n$ always has two
unitary irreducible representations (UIR's) of dimension 1 - the trivial representation and the antisymmetric (sign)
representation.  In ordinary three dimensional space, these correspond to Bose-Einstein and Fermi-Dirac statistics
respectively.  Fractional statistics could in theory arise in three dimensions, but would have to correspond to a higher
dimensional representation of the symmetric group.  However, as explained by Haag in~\cite{haag}, any physical theory including
fermions obeying parastatistics in three dimensions can be modeled as a theory including only regular fermions by including
some hidden variable, without changing any of the resulting physics.

\subsubsection{Two Dimensions}

\textbf{\textit{Indistinguishable Particles: Braid Group}}
On the other hand, in two dimensions the fundamental group of the configuration space of a system of $n$ identical particles
is given by the more general braid group:
$$\pi_1(C_n(\R^2)) \simeq B_n(\R^2) = \langle \sigma_i,  1 \leq i \leq n-1 | \sigma_i \sigma_{i+1} \sigma_i = \sigma_{i+1} \sigma_i \sigma_{i+1}, $$
$$ \sigma_i \sigma_j = \sigma_j \sigma_i \forall |i-j| \geq 2 \rangle .$$
The $\sigma_i$ are defined physically by interchanging the $i^{th}$ and $(i+1)^{st}$ particles counterclockwise.  In
contrast to the symmetric group, not all braid group generators satisfy $\sigma_i^2 = Id$.  The monodromy operator/$ R^2$ matrix is defined as the square of the braiding operator/R matrix.

Intuitively, two dimensional
space allows for more possibilities because particles are topologically pointlike.  Adding a particle to a 2D surface is
like poking a hole in the surface, so loops around the hole can no longer be continuously deformed to a single point.
(Note that our motivational Aharonov-Bohm effect example simulates 2D as the solenoid is infinite in length, so only 2 of
the dimensions really matter.  This corresponds to the fact that the loops around a line are characterized by the
fundamental group (first homotopy group), $\pi_1(\R^2 \setminus line) = \pi_1(S^1 \times \R) = \Z $.)

\textbf{\textit{Truncated Braid Group}}
The braid group as defined is infinite and torsion-free (all elements of the group have infinite order).  For theories with a finite symmetry group, it is actually the truncated braid group $B(m,n)$, that governs the interaction of particles.  These are the braid groups on $n$ elements with
the additional condition that $\sigma^m=1$ for some finite $m$ ($m$ is taken to be minimal.  $m=2$ corresponds to the case of
$S_n$, so the truncated braid group is indeed a generalization of the symmetric groups.) When $\sigma$ acts on physical
particles, the Hamiltonian transforms as a representation of $\sigma$ (the braid operator), which we call the $R$ matrix.
(More on this topic will be discussed in the next section.)

\textbf{\textit{Distinguishable Particles: Colored (Pure) Braid Group}}
If the particles are distinguishable, the configuration space is given by $C((\R^2)^n - D)$.  The colored (pure) braid group $P(m,n)$, for $n$ even, describes the fundamental group of the configuration space of the underlying manifold.  (Intuitively, only monodromy operators make sense since after each exchange the two particles have to be back in their original positions.)  Formally, the pure braid group $P(m,n)$ is the kernel of the homomorphism $\phi: B_n \rightarrow S_n$ mapping each braid to the associated
corresponding permutation.  

\textbf{\textit{Partially Colored Braid Group}}  If the system consists of a combination of distinguishable particles and indistinguishable particles, then the fundamental group is the partially colored braid group.

\section{Spin-Statistics Connection}

\subsection{The Abelian Case}

The topological approach to proving the spin-statistics
connection (~\cite{prop})
that depends only on the existence of antiparticles and Lorentz invariance is useful when considering the spins of particles in 2 dimensions.
(This proof was originally proposed by Finkelstein and Rubinstein in the case of three spatial dimensions, then adapted to the two dimensional case by
Propitius de Wild.~\cite{prop})  Consider the process in which two particle/antiparticle pairs are created from the vacuum, the
particles are interchanged, then the new pairs annihilate.  The associated spacetime ribbon can be deformed via Lorentz transformations into the ribbon
for rotation of a single one of the particles by an angle of $2 \pi$ about its center, as can be seen from diagram [2].  
Thus from the topological perspective, these two processes are seen to be physically equivalent.

\begin{figure}[t]
\centerline{\includegraphics[width=15cm, height=10cm]{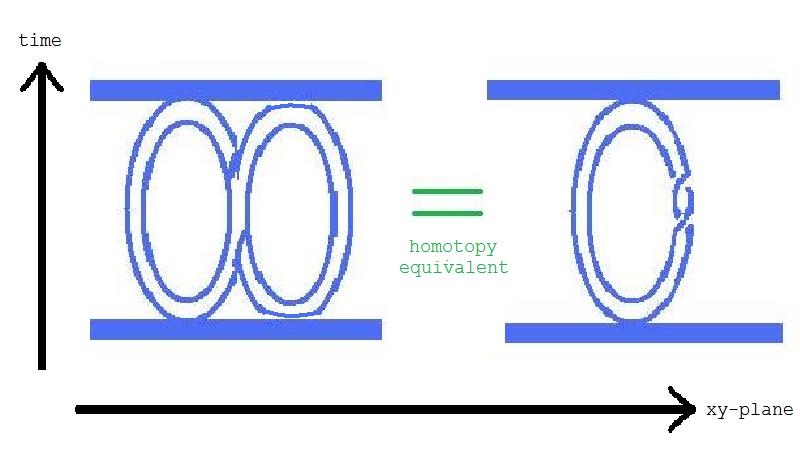}}
\caption[Spin-Statistics Connection for an Abelian Group]{Note that the effect of interchanging two particles in two separate particle/antiparticle pairs can be continuously deformed into the process under which one of the particles is rotated by an angle of $2 \pi$ about its center.  Thus the two effects should be topologically the same.}
 \label{fig:SpinStatistics}
\end{figure}

Thus the statistical phase factor $e^{i \theta}$ picked up by exchanging the two particles must in
fact be equal to the spin factor $e^{2\pi i s}$ obtained by rotating a particle of spin $s$.
Note that the procedure in the above proof is independent of whether the particles or the antiparticles are interchanged
(the same orientation results either way).  Thus as an additional consequence we obtain that the spin of the antiparticle is
equal to that of the corresponding particle.

\subsection{The Non-Abelian Case: Generalized Spin-Statistics Connection}

When the group is non-abelian, the canonical spin statistics connection must be generalized as the topological equivalence in Fig.[3] 
 shows.  Specifically, when a composite particle C (the result of fusion of A and B) is rotated, not only is the phase of the composite picked up, but each of the constituents A and B are also rotated.  Thus both A and B must both be rotated back in the opposite direction, each picking up a phase.

$$K_{\alpha \beta \gamma}^{ABC} R^2 = e^{2 \pi i (s_{(C, \gamma)}-s_{(A, \alpha)}-s_{(B, \beta)})} K^{ABC}_{\alpha \beta \gamma}$$

\begin{figure}[t]
\centerline{\includegraphics[width=15cm, height=10cm]{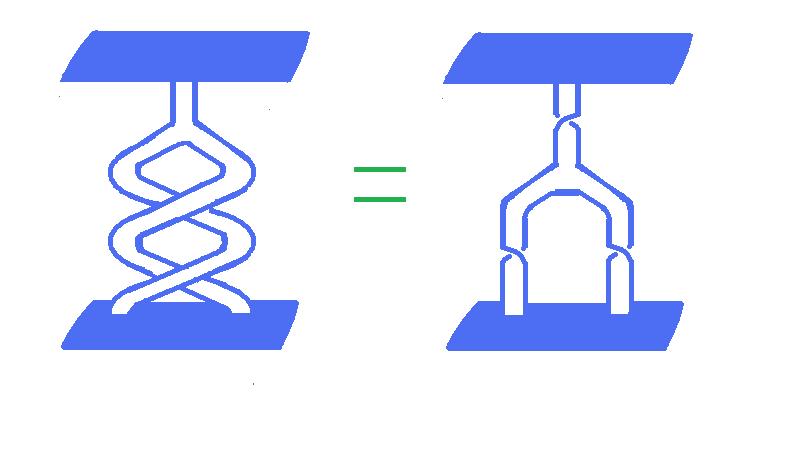}}
\caption[Generalized Spin-Statistics Connection for an Abelian Group]{The Generalized Spin-Statistics Connection applies to two particle fusion in non-abelian theories.  On the left, fusion is performed after a monodromy operation.  This can be continuously deformed via Lorentz transformations to the diagram on the right, in which the particles are each twisted by a $-2\pi$ rotation before being fused, then the fusion product is twisted by $2\pi$.}
 \label{fig:GenSpinStatistics}
\end{figure}

When the two particles are the same, a square root version of the generalized Spin-Statistics Connection is satisfied:

$$ K_{\alpha \alpha \gamma}^{AAC} R = \epsilon e^{\pi i (s_{(C,\gamma)}-2s_{(A, \alpha)})} K_{\alpha \alpha \gamma}^{AAC} $$

\section{Determining the Particles from the Hamiltonian}

\subsection{Abelian}

Consider the following action for a minimally coupled scalar field with an Abelian symmetry group:

$$ S = S_{YMH} + S_{matter} = \int d^3 x (-\dfrac{1}{4} F_{\mu \nu} F^{\mu \nu} + (D^{\mu} \Phi)^* D_{\mu}\Phi - V(|\Phi|) - j^{\mu}A_{\mu} )    $$ 
where the covariant derivative is given by $D_{\rho} \Phi = (\partial_{\rho} + i N e A_{\rho}) \Phi $ (i.e. the Higgs field has charge $Ne$ with respect to the compact $U(1) $ symmetry).  The last term represents the contribution from the matter fields while the others are the Yang-Mills-Higgs contribution.  Terms in the Lagrangian density that can be written as a total derivative are not included as usually the terms cancel at the boundaries.  (This is not always the case.  In particular, Chern-Simons theories contain an additional term which is a total derivative but the boundary conditions still contribute precisely because there is a non-trivial topological winding at the boundary as in the case, for instance, of axions.  There is also a description of the Fractional Quantum Hall effect in terms of Chern-Simons theories.  However, in this paper, the emphasis is on topological effects due solely to the ordinary Lagrangian terms.)

\subsection{Non-Abelian Cases}
In the case of the nonabelian Higgs field, the field tensor must be given an extra index $a$.
$$S_{YMH} = \int d^3x (-\dfrac{1}{4} F^{a \mu \nu} F^a_{\mu \nu} + (D^{\mu} \Phi)^\dag \cdot D_{ \mu } \Phi - V(\Phi)) $$

where the covariant derivative is given, for instance, in the adjoint representation of the nonabelian group by $ (D_{\mu} \Phi )_a = \partial_{\mu} \phi_a+ g f^{abc} A_{\mu}^b \Phi_c $.  Explicitly, the field strength is $F^a_{\mu \nu} = \partial_{\mu} A^a_{\nu} - \partial_{\nu} A^a_{\mu} - g f^{abc} A^b_{\mu} A^c_{\nu}$.

\subsection{Determining the Particles}

Note that there is spin-charge separation in the system as the flux and the charge can be measured independently (and there are both charged particles with no magnetic flux and magnetic fluxes with no electric charge).
A specific particle type is determined by a conjugacy class A of the symmetry group H and a representation $\alpha$ of the
centralizer of a representative of that conjugacy class.  (The centralizers of each element of the conjugacy class are isomorphic, so the representation thus defined is independent of the particular representative chosen.)  In particular, label the elements of each conjugacy class
$^A C = \{^A h_1, ^A h_2, ..., ^A h_k \}$.  Let $^A N$ denote the subgroup of H that is the centralizer for $^A h_1$.  Choose a set of
representatives of $H/^A N = \{^{\alpha}v_1,...,^{\alpha}v_k\} $.  The states  $$ \{ |^A h_i, ^{\alpha}v_j \rangle \}$$ then
constitute a basis for the internal Hilbert space $V_{\alpha}^A$.

A bit more formally, given the flux values take values in a group H, then quantum mechanically we should consider complex combinations $\C[H]$.  The electric charge takes values in the dual $\textit{F}(H)$.  
The particles thus correspond to representations of Drinfeld's quantum double construction applied to the Abelian algebra $\textit{F}(H)$ of functions on a finite group $H$.  Generators are given by combinations $\{P_h g\}_{h,g \in H}$, where $P_h$ is the operator projecting the
flux $h \in H$ and $g \in H$ corresponds to the global symmetry transformation.  The operators $P_h$ are projections as $P_h P_{h'} = \delta_{h, h'}P_h$.  We demonstrated earlier that the action of the
global symmetry transformations does not commute with the flux projection operators, and in fact: $ g P_h = P_{g h g^{-1}} g$
due to flux metamorphosis.  Combining these two relations we obtain the multiplication rule for two elements of the quantum
double: $P_h g \cdot P_{h'}g' = \delta_{h, g h' g^{-1} P_h g g} $.  The particles permitted by a discrete gauge theory are given by unitary irreducible representations of the quantum double.  Note that since the quantum double is an algebra, not a group, representations $\Pi$ here are defined not only to respect the multiplicative structure, but are also required to be linear, i.e. addition and scalar multiplication are preserved.  Each unitary irreducible representation of $D(H)$ is labeled uniquely by fixing a conjugacy class of the underlying group
H and specifying the representation of the centralizer of a representative element in that conjugacy class.  The algebra is semisimple, i.e. it can be decomposed into a direct sum.  Thus the sum of squares of internal Hilbert spaces = the dimension of the quantum double. (This will be used as a consistency check to confirm the explicit dimensions obtained in future sections.)

\subsection{Changing Basis: The S-Matrix}

The S-matrix implements the change of basis between the given basis of the fusion algebra and the diagonal basis.  (Such a
matrix S must exist since the algebra is commutative and associative.)  Explicitly, the S matrix can be written as:

$$S_{\alpha \beta}^{AB} := \dfrac{1}{|H|} tr \textrm{ }R^{-2}{ }^{AB}_{\alpha \beta} $$
$$ = \dfrac{1}{|H|} \sum_{^Ah_i \in ^AC, ^Bh_j \in ^BC: [^Ah_i, ^Bh_j]=e} tr(\alpha(^Ax_i^{-1} { }^Bh_j ^Ax_i))^* tr (\beta(^Bx_j^{-1}{ }^Ah_i ^Bx_j))^* $$

In the appendices, a MAGMA program that computes the S matrix values is included.

\subsection{Spin and the T-Matrix}

The diagonal T-matrix records the information about the spins.  Explicitly, the T matrix is given by:

$$ T_{\alpha \beta}^{AB} := \delta_{\alpha,\beta} \delta^{A,B} e^{2\pi i s_{(A, \alpha)}} $$

Within each internal Hilbert space, all particles carry the same spin.  Furthermore, the only particles which have nontrivial spin are dyons with both
nontrivial flux and nontrivial charge.  Note that in the construction for the quantum double, by definition the conjugacy class A representative element $^A h_1$ commutes with all elements in the centralizer $^A N$, so Schur's lemma yields:

$$\alpha(^A h_1) = e^{2 \pi i s_{(A, \alpha)}} \textbf{1}_{\alpha}$$

for every irreducible representation $\alpha$.  This formula is implemented in the MAGMA program to yield the values of the spin for the various anyonic particles.  While this does not distinguish between spin
$s=0$ and $s=1$, it does the best that is possible in any superselection sector as it is always possible to add orbital
angular momentum.  In the specific groups discussed in the next section, a number of exotic spins are discovered.

\subsection{Magnetic-Electric Charge Duality}

  The symmetry transformations of the second factor ($\C(H)$) in the quantum double construction relate $D(H)$ to the
  electric charge.  Similarly, the first part $F(H)$ of $D(H)$ yields the magnetic fluxes.  Note that these are in fact dual
  to each other, as $F(H)\simeq \C H^* $.  The magnetic-electric charge duality is implemented by the $SL(2, \Z)$ action generated by the S and T matrices.  The action of the $S$ matrix has an interpretation as a Fourier transform.~\cite{koorn}

\subsubsection{Fusion} Similar to regular quantum field theory of the collision of two particles in three dimensions, the fusion of two
particles can lead to multiple possible outcomes.  The fusion rules are the analog in this two dimensional setting of the
selection rule for addition of angular momenta $J_1$, $J_2$ in quantum mechanics which permits values between $-|J_2-J_1|$
and $J_2+J_1$ as feasible outcomes for the combined system.  The fusion coefficients indicate how many different ways there
are to obtain that outcome.  (However, this still does not tell us the probabilities of obtaining the specific outcomes, so
it is not quite the analog of the Clebsch-Gordon coefficients.)

The fusion coefficients are encoded in the comultiplication of the dual $D(H)^*$.  Explicitly,

$$ N^{AB \gamma}_{\alpha \beta C} = \sum_{D,\delta} \dfrac{S_{\alpha \delta}^{AD} S_{\beta \delta}^{BD} (S^*)_{\gamma \delta}^{CD}}{S_{0 \delta}^{eD}}$$

where as before the modular S matrix is given by: $$S_{\alpha \beta}^{AB} := \dfrac{1}{|H|} tr \textrm{ }R^{-2}{ }^{AB}_{\alpha \beta} $$
$$ = \dfrac{1}{|H|} \sum_{^Ah_i \in ^AC, ^Bh_j \in ^BC: [^Ah_i, ^Bh_j]=e} tr(\alpha(^Ax_i^{-1} { }^Bh_j ^Ax_i))^* tr (\beta(^Bx_j^{-1}{ } ^Ah_i ^Bx_j))^* $$

\subsubsection{Antiparticles}

An antiparticle is defined to be the representation which when fused with the original particle yields the identity as one
possible outcome.  It turns out that the antiparticle is unique, and the charge conjugation operator $C=S^2$ takes a particle to its antiparticle.
Note that since the S matrix commutes with the T matrix (which encodes the spin), the spin of a particle is the same as
that of the corresponding antiparticle.


\subsection{Alice Physics and Cheshire Charges in Non-Abelian Gauge Theories}

Cheshire charge is a non-localizable form of charge that is carried by a configuration of particles.  Globally, charge must be conserved.  Thus, a special type of entanglement is observed if the particles with Cheshire charge are kept far apart from one another so that charge cannot be transferred by local interactions.  This entanglement opens up the possibility of topological quantum computation using anyons.  In particular, consider a collection of particle/antiparticle pairs.  Each local particle/antiparticle pair has trivial total flux and charge - If kept sufficiently separated, the environment cannot distinguish between states with long range quantum numbers the same as those given by the ground state.  However, the configuration still has nontrivial Cheshire charge, so the state cannot annihilate.  In other words, since the environment acts locally on the system, it cannot effect the non-local Cheshire charge carried by the configuration.  Thus the system is protected against decoherence.

\section{Summary}
The topology of the configuration space restricts the possible symmetry groups of the Hamiltonian.  In turn, every term in
the Hamiltonian must transform as a singlet under all the representations of the symmetry group so this restricts the types
of terms that can appear in the Hamiltonian.  In (2+1)-dimensional systems, the truncated braid group must be used instead of the symmetric group.  Consistent braiding and fusion then impose conditions which must be solved by any particles in the theory.  Particles have two degrees of freedom given by the magnetic flux and electric charge.  In (2+1)-dimensional theories based on a finite symmetry group, representations of the quantum double of the symmetry group determine the kinds of particles that can possibly exist in the given theory.  The fusion rules are then encoded by the comultiplication of the dual.

\section{Explicit Derivations of Particles and Fusion Rules for Special Physical Theories}
\label{exs}


\subsection{Abelian vs. Non-Abelian: Physical Origin and Computational Potential}

In Abelian groups, the fusion rules obtained by the quantum double formalism are deterministic
as only one product can result from fusion of two particles.  However, in non-Abelian Groups,
the fusion rules obtained for at least two input particles include multiple possible output particles so the product is not determined
solely by the symmetry group.  (Letting $N^C_{AB}$ be the number of ways of obtaining particle $C$ when the particles $A$ and $B$ are fused, mathematically this condition is just $ \sum_C N^C_{AB} > 1 $ for some particles $A,B$.)  To obtain a deterministic result, the more detailed
dynamics of the specific interaction have to be taken into account.

From another perspective, non-Abelian groups lead to multidimensional Hilbert Spaces so there is
more wiggle room for changes in the state (i.e. the operators can be noncommutative in a multidimensional space).  This is the key behind the computational power of non-Abelian groups in quantum computation.

Abelian groups are severely limited in their applications to quantum computation by the fact that they can only be used to store information
(not to process it).  However, Abelian groups nonetheless underlie some physical systems and are used in Kitaev's Toric Code
for storage~\cite{kitaevToric} and are included here for completeness.  Specifically, superconductivity in a material corresponds to SSB of the electromagnetic $U(1)$ gauge field -
the standard condensed matter Cooper Pair, for example, arises from the SSB of the electromagnetic $U(1)$ to $\Z/N\Z $
with $N=2$.

\subsection{Applying the Quantum Double Formalism to Abelian Finite Groups $\Z/N\Z$} 
Finite abelian groups are the simplest case that has been worked out - computer algebra is not required.~\cite{prop}  Here we include the results for completeness and so we can analyze the tables for Majorana particles.  
Consider the effect of spontaneous symmetry breaking of an Abelian $U(1)$ gauge group (or the nonabelian $SO(3)$, $SU(2)$, or $SU(3)$ gauge groups) down to the finite abelian subgroup $\Z/N\Z$.  Screening precludes electromagnetic effects between two particles from influencing the
long-distance interactions.  However, the topological charges persist.  For these Abelian theories, the monodromy operator is just a phase.

\subsubsection{Particles}

The spectrum in $\Z/N\Z$ theories consists of pure charges, pure fluxes, and dyons produced by fusing these charges and fluxes.
Let
$$ | a \rangle | n \rangle = |a, n \rangle $$
denote a flux $a$ ($\in \Z/N\Z$), charge $n$ ($\in \Z/N\Z$) particle.

\subsubsection{Braiding and Fusion}

The braiding and fusion rules for $\Z/N\Z$ theories are particularly simple.  Braiding two identical particles, one encounters
the Dirac string of the other, which introduces a phase:

$$ R | a , n \rangle | a , n \rangle = e^{\frac{2 \pi i}{N} (n a) } | a , n \rangle  |a, n \rangle $$

For non-identical particles, only the monodromy operator makes sense:
$$ R^2 | a , n \rangle | a' , n' \rangle = e^{\frac{2 \pi i}{N} (n a' +n' a) } | a , n \rangle  |a', n' \rangle $$

The fusion rules for this theory amount to summing the charges and fluxes mod N:

$$ | a , n \rangle \times | a' , n' \rangle = | [a+a'] , [n+n'] \rangle$$

Instantons in effect perform the modulo N summation.
The charge conjugation operator takes each charge, flux combination to the additive inverse charge, flux mod N.  This in fact
corresponds to the antiparticle.  Note that the only $\Z/N\Z$ gauge theories which contain a Majorana particle (i.e. a particle that is its own antiparticle) are those for which $N$ is even.  In that case, the only Majorana particle has both flux and charge $\frac{N}{2}$.

%
%
%
%

\subsection{Applying the Quantum Double Formalism to Non-Abelian Finite Groups}

\subsubsection{Finite Subgroups of $SO(3)$ and $SU(2)$}

Non-abelian quantum doubles arise from SSB of non-abelian groups.  $SO(3)$ has five finite subgroup types, all arising from symmetries of regular polygons and polyhedra  -
\begin{enumerate}
\item the cyclic groups $\Z/N\Z$,
\item the dihedral groups $D_n$,
\item the symmetries of the tetrahedron $T=A_4$,
\item the symmetries of the cube or octahedron $S_4$, and
\item the rotational symmetries of the icosahedron $A_5$.
\end{enumerate}
The abelian quantum doubles $\Z/N\Z$ were tabulated above.  In this section the non-abelian doubles $S_3=D_3$ and $A_5$ are considered, as these are liable to be relevant for quantum computation.
The tabulation of the quantum doubles of the next dihedral group and all non-dihedral finite subgroups of $SO(3)$ will be completed in the appendices.  




These same subgroups (up to a semidirect product with $\Z/2\Z$) also describe all the finite subgroups of $SU(2)$.  
Specifically, all the finite subgroups of the double cover $SU(2)$ arise from liftings from $SO(3)$ under the $2:1$ projection map $\pi: SU(2) \rightarrow SO(3)$.  Observe that the only element of order 2 in $SU(2)$ is $-I$, which generates the kernel.  Recall that all even order groups, and only the even order groups, have an element of order 2.  Thus the finite subgroups of $SU(2)$ of even order must correspond to the preimages under $\pi$ of the finite subgroups of $SO(3)$ and any finite subgroup of odd order must be isomorphic to an odd order subgroup of $SO(3)$.  Each finite subgroup $H$ of $SO(3)$ lifts to $H \ltimes \Z/2\Z $ (the subgroup semidirect product $\Z/2\Z$) in $SU(2)=Spin(3)$.  These are known as the ``binary" versions of the $SO(3)$ subgroups.  
The odd order subgroups of $SO(3)$ (and thus $SU(2)$) are the cyclic groups $\Z/N\Z$ with $N$ odd.



Note also that $SO(3)$ is a subgroup of $SU(3)$.  Thus all the finite subgroups enumerated here could also be formed from SSB of $SU(3)$.  The results for the remaining finite subgroups of $SU(3)$ will be presented elsewhere.


\subsubsection{$S_3$}
The conjugacy classes of the symmetric groups are determined by the cycle decomposition.  In particular, for $S_3$, there are
3 conjugacy classes with representative elements (), (12), (123) - These correspond to fluxes in the spontaneously broken gauge theory.
The centralizers of a representative element of each conjugacy class are given by: $S_3$, $\Z/{2\Z}$, and $\Z/{3\Z}$, respectively.  The
possible charges are given by representations of the centralizer corresponding to the particle's flux.

The pure charge sector corresponds to flux () and representations of the centralizer $S_3$.  Since $S_3$ has three conjugacy classes,
there are three irreducible representations of $S_3$, with dimensions 1,1, and 2 (the unique unordered solution $\{n_1,n_2,n_3\}$ to the equation $n_1^2+n_2^2+n_3^2 = 6 = |S_3|$).  These are given by the identity representation, the sign representation, and a 2 dimensional representation.

For $S_3$, the particles with nontrivial flux can have charges that are representations of $\Z/{2\Z}$ and $\Z/{3\Z}$ respectively.
These last two centralizer groups are Abelian - the corresponding representations are thus just roots of unity.

\begin{itemize}
\item Flux: Identity Element (): In this case, the charges are given by representations of the centralizer $S_3$:

\begin{center}
\begin{tabular}[t]{l|c|c|c}
\multicolumn{4}{c}{Representations of $S_3$} \\ \hline
Repn&()&(12)&(123)\\\hline
$\chi_1$&1&1&1\\
$\chi_2$&1&-1&1\\
$\chi_3$&2&0&-1\\\hline
\end{tabular}
\end{center}

\item Flux: Two-cycle Conjugacy Class (Representative Element: (12)): In this case, the charges are given by representations of the Centralizer $\Z/2\Z$.
\begin{center}
\begin{tabular}[t]{l|c|c}
\multicolumn{3}{c}{Representations of $\Z/2\Z$}\\\hline
Repn&()&(12)\\\hline
$\rho_1$ &1&1\\
$\rho_2$ &1&-1\\\hline
\end{tabular}
\end{center}

\item Flux: Three-cycle Conjugacy Class (Representative Element: (123)):  In this case, the charges are given by representations of the Centralizer $\Z/3\Z$.
\begin{center}
\begin{tabular}[t]{l|c|c|c}
\multicolumn{4}{c}{Representations of $\Z/3\Z$} \\ \hline
Repn&()&(123)&(132)\\\hline
$\tilde\rho_1$ &1&1&1\\
$\tilde\rho_2$ &1& $e^{2\pi i/3}$ & $e^{4\pi i/3}$ \\
$\tilde\rho_3$ &1& $e^{4\pi i/3}$ & $e^{2\pi i/3}$ \\ \hline
\end{tabular}
\end{center}

\end{itemize}

In summary then, there are 8 different types of particles, given by:

\begin{center}
  \begin{tabular}{ c | c | c | c | c | c}
    Particle & Flux & Charge & Spin ($e^{2\pi i s_{(A, \alpha)}}$) & Repn Dim & Dim Internal Hilbert Space \\    \hline
   A & () & $\chi_1$ & 0 & 1 & 1  \\
   B & () & $\chi_2$ & 0 & 1 & 1\\
   C & () & $\chi_3$ & 0 & 2 & 2\\ 
   D & (12) & $\rho_1$ & 0 & 1 & 3\\ 
   E & (12) & $\rho_2$ & $\frac{1}{2}$ & 1 & 3 \\
   F & (123) & $\tilde\rho_1$ & 0 & 1 & 2 \\
   G & (123) & $\tilde\rho_2$ & $\frac{1}{3}$ & 1 & 2\\
   H & (123) & $\tilde\rho_3$ & $\frac{2}{3}$ & 1 & 2\\

  \end{tabular}
\end{center}

where the Spin $s$ is chosen in the range $0 \leq s <1$.

Note that particles E, G, and H all have fractional spin - particle E is a regular fermion, but G and H have spin
$\frac{1}{3}$ and $\frac{2}{3}$ respectively, so this is an explicit demonstration of fractional statistics.
As a check to this table, we do indeed have the equivalence that the sum of the dimensions of the internal Hilbert spaces
is $1^2+1^2+2^2+2*3^2+3*2^2=36=|S_3|^2$, the dimension of the quantum double.

The monodromy operator is given by the square of the braiding operator.

\underline{\textbf{Fusion Rules}}

\begin{figure}[t]
\centerline{\includegraphics[width=13.5cm, height=8cm]{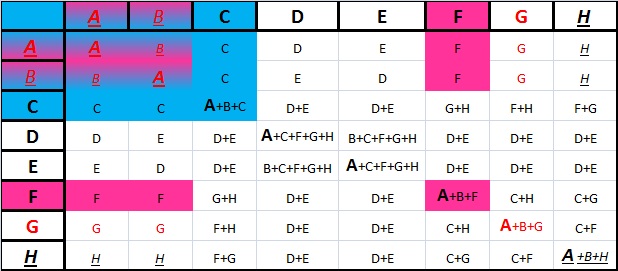}}
\caption[$S_3$]{Fusion Rules for $S_3$}
\label{fig:S3}
\end{figure}

Note that in the table for $S_3$, every particle is its own antiparticle.  Also, there are many subsystems which are 
closed under fusion.  In particular, the subsystems \begin{verbatim}{A,B}, {A,B,C}, {A,B,G}, {A,B,F}, {A,B,H}, {A,B,F,G}, {A,B,F,H}, {A,B,G,H}, \end{verbatim}  \begin{verbatim}{A,B,F,G,H} and {A,B,C,F,G,H} \end{verbatim}
are all closed under fusion.  Thus if not all the particles were initially present, then they need not arise by either fusion or braiding.  However, as soon as D or E particles are present in the system, with some probability all of the other particles will appear.  In addition, quantum mechanically virtual pair production can occur so all the particles appear with some probability, if only for a short time, regardless of the initial particle content.

Note that the $\{ A, B, C\}$ subsystem may be physically natural in a theory in which the pure electric charges A, B and C are much lighter than the rest of the particles so we can integrate the others out.  This is potentially reasonable because magnetic fluxes tend to be much heavier than their trivial flux counterparts.  The three-particle subsystems will be analyzed further in the section on quantum computation.

Finally, there are a number of unexplained symmetries in the fusion table that a priori would not necessarily be expected based on the $S_3$ symmetry group of the underlying theory.  In particular, the fusion table is invariant under any permutation of the labels C, F, G, and H.  Physically, however, the C particle tends to have a much smaller mass than the F, G, and H particles since it is a pure electric charge.

\underline{\textbf{$S_3$ Symmetry Group Implementations}} 

Note that $S_3$ is isomorphic to $C_{3v}$, the symmetry group of a honeycomb lattice as observed naturally in, for instance, graphene.  Thus it is perhaps reasonable to expect to find an $S_3$ system conducive to topological gauge theories in nature.  Alternately, $S_3 = D_3 = \Z/3\Z \ltimes \Z/2\Z $, so $S_3$ could potentially arise from SSB in chiral p-wave superconductors.

$O(2) = U(1) \ltimes \Z/2\Z$ broken down to $S_3=<R_{120}, flip>$ where $R_{\omega}$ denotes a counterclockwise rotation by an angle $\omega$.

In addition, hexagonal optical lattices can be manufactured using cold atoms.  L. Duan, E. Demler, and M. Lukin proposed an implementation of Kitaev's code on the honeycomb lattice.~\cite{luk}  Finally, a method for creating $S_3$ anyons in Josephson junction arrays has been suggested.~\cite{JJ}

\subsubsection{$A_5$}
The conjugacy classes of $A_5$ are given by (), (12)(34), (123), (12345), (12354).  The centralizers of a representative element of
each of these conjugacy classes are given by $A_5$, $\Z/2\Z \times \Z/2\Z$ (= group generated by (12)(34) and (13)(24)), $\Z/3\Z$
(= group generated by (123)), $\Z/5\Z$ (= group generated by (12345)), and $\Z/5\Z$ (= group generated by (12354)) respectively.
The dimensions of the internal Hilbert spaces of these fluxes are given by the number of elements in the conjugacy class (a basis is
given by linear combinations of elements in this conjugacy class).  These dimensions are thus: 1, 15, 20, 12, 12  respectively.
(The order of the conjugacy class times the order of the centralizer equals order of the group, as it must by elementary group theory.)

First consider the pure charge sector.  These are elements that arise from irreducible representations of $A_5$.  Since there are 5
conjugacy classes, there are 5 irreducible representations of $A_5$.  The dimensions of the irreducible representations are
1, 3, 3, 4, 5.  Note that the sum of the squares of the dimensions of these representations is equal to $|A_5|=60$ as it
should.

The character table for $A_5$ is thus:

\begin{center}
  \begin{tabular}{ c | c | c | c | c | c }
    Rep & () & (12)(34) & (123) & (12345) & (12354) \\    \hline
    $\tau_1$ & 1 & 1 & 1 & 1 & 1 \\
    $\tau_2$ & 3 & -1 & 0 & A & *A \\
    $\tau_3$ & 3 & -1 & 0 & *A & A \\ 
    $\tau_4$ & 4 & 0 & 1 & -1 & -1 \\ 
    $\tau_5$ & 5 & 1 & -1 & 0 & 0 \\ 
  \end{tabular}
\end{center}

where $A=-e^{\frac{4\pi i}{5}}-e^{\frac{2\pi i}{5}}$

Now consider the dyons (elements of both flux and charge) - There are 4 possible fluxes other than the identity, labeled by
the conjugacy classes of (12)(34), (123), (12345), and (12354) respectively.  The possible electric charges for a particle
with a flux corresponding to the conjugacy class of (12)(34) are given by representations of the centralizer.  Specifically,
the centralizer is the Klein-4 Group: $\Z/2\Z \times \Z/2\Z$, with representations:

\begin{center}
  \begin{tabular}{ c | c | c | c | c }
    Rep & e & (12)(34) & (13)(24) & (14)(23) \\    \hline
    $\eta_1$ & 1 & 1 & 1 & 1 \\
    $\eta_2$ & 1 & -1 & -1 & 1 \\
    $\eta_3$ & 1 & -1 & 1 & -1 \\ 
    $\eta_4$ & 1 & 1 & -1 & -1 \\ 
  \end{tabular}
\end{center}

The possible electric charges associated with the flux corresponding to the
conjugacy class of (123) are given by representations of the
centralizer $\Z/3\Z$ of this flux.  Explicitly:

\begin{center}
  \begin{tabular}{ c | c | c | c }
    Rep & e & (123) & (132)   \\    \hline
    $\tilde\rho_1$ & 1 & 1 & 1  \\
    $\tilde\rho_2$ & 1 & $e^{2\pi i/3}$ & $e^{4\pi i/3}$  \\
    $\tilde\rho_3$ & 1 & $e^{4 \pi i/3}$ & $e^{2 \pi i/3}$  \\ 
  \end{tabular}
\end{center}

Finally, consider the fluxes corresponding to the 5 cycles, (12345) and (12354).  The electric charges in both cases are
given by representations of $\Z/ 5\Z $, so the representation table will be listed only once:

\begin{center}
  \begin{tabular}{ c | c | c | c | c | c}
    Rep & e & (12345) & (13524) & (14253) & (15432)\\    \hline
    $\tilde{\tilde{\tilde\rho}}_1$ & 1 & 1 & 1 & 1 & 1 \\
    $\tilde{\tilde{\tilde\rho}}_2$ & 1 & $e^{2\pi i/5}$ & $e^{4 \pi i/5}$ & $e^{6\pi i/5}$ & $e^{8 \pi i/5}$\\
    $\tilde{\tilde{\tilde\rho}}_3$ & 1 & $e^{4\pi i/5}$ & $e^{3\pi i/5}$ & $e^{2\pi i/5}$ & $e^{6\pi i/5}$ \\ 
    $\tilde{\tilde{\tilde\rho}}_4$ & 1 & $e^{6\pi i/5}$ & $e^{2 \pi i/5}$ & $e^{8 \pi i/5}$ & $e^{4 \pi i/5}$ \\ 
    $\tilde{\tilde{\tilde\rho}}_5$ & 1 & $e^{8\pi i/5}$ & $e^{6\pi i/5}$ & $e^{4\pi i/5}$ & $e^{2\pi i/5}$
  \end{tabular}
\end{center}


The particles are thus given by:

\begin{center}
  \begin{tabular}{ c | c | c | c | c | c}
    Particle & Flux & Charge & Spin & Dim Charge Repn & Dim Internal Hilbert Space \\    \hline
   A & () & $\tau_1$ & 0 & 1 & 1  \\
   B & () & $\tau_2$ & 0 & 3 & 3\\
   C & () & $\tau_3$ & 0 & 3 & 3\\ 
   D & () & $\tau_4$ & 0 & 4 & 4\\ 
   E & () & $\tau_5$ & 0 & 5 & 5 \\
   F & (12)(34) & $\eta_1$ & 0 & 1 & 15 \\
   G & (12)(34) & $\eta_2$ & 0 & 1 & 15\\
   H & (12)(34) & $\eta_3$ & $\frac{1}{2}$ & 1 & 15\\
   I & (12)(34) & $\eta_4$ & $\frac{1}{2}$ & 1 & 15\\
   J & (123) & $\tilde\rho_1$ & 0 & 1 & 20 \\
   K & (123) & $\tilde\rho_2$ & $\frac{1}{3}$ & 1 & 20 \\
   L & (123) & $\tilde\rho_3$ & $\frac{2}{3}$ & 1 & 20 \\
   M & (12345) & $\tilde{\tilde{\tilde\rho}}_1$ & 0 & 1 & 12\\
   N & (12345) & $\tilde{\tilde{\tilde\rho}}_2$ & $\frac{1}{5}$ & 1 & 12\\
   O & (12345) & $\tilde{\tilde{\tilde\rho}}_3$ & $\frac{2}{5}$ & 1 & 12\\
   P & (12345) & $\tilde{\tilde{\tilde\rho}}_4$ & $\frac{3}{5}$ & 1 & 12\\
   Q & (12345) & $\tilde{\tilde{\tilde\rho}}_5$ & $\frac{4}{5}$ & 1 & 12\\
   R & (12354) & $\tilde{\tilde{\tilde\rho}}_1$ & 0 & 1 & 12\\
   S & (12354) & $\tilde{\tilde{\tilde\rho}}_2$ & $\frac{1}{5}$ & 1 & 12\\
   T & (12354) & $\tilde{\tilde{\tilde\rho}}_3$ & $\frac{2}{5}$ & 1 & 12\\
   U & (12354) & $\tilde{\tilde{\tilde\rho}}_4$ & $\frac{3}{5}$ & 1 & 12\\
   V & (12354) & $\tilde{\tilde{\tilde\rho}}_5$ & $\frac{4}{5}$ & 1 & 12\\
  \end{tabular}
\end{center}


In summary then, there are 22 different types of particles.  As a check, since the algebra is semisimple, the sum of the squares of the dimensions of the
internal Hilbert spaces for the particles should equal $|A_5|^2 $, the order of the quantum double.  Now, the dimension of
the internal Hilbert space is given by the product of the number of elements in the specific conjugacy class with the
dimension of the irreducible representation.  Thus the dimension is  $1^2+3^2+3^2+4^2+5^2+5^2+4*(15*1)^2+3*(20*1)^2+2*5*(12*1)^2 = 3600 $.
This does indeed equal $|A_5|^2 = 60^2 = 3600$, as desired.

\subsubsection{Fusion Rules:}

The fusion rules for $A_5$ can be found in the following pages.  Once again every particle is its own antiparticle.  Note that, unlike in the $S_3$ case, there are multiple ways of obtaining some of the products - For instance, E and F can fuse to produce E in two distinct ways.

\begin{figure}[t]
\centerline{\includegraphics[width=19cm, height=9cm]{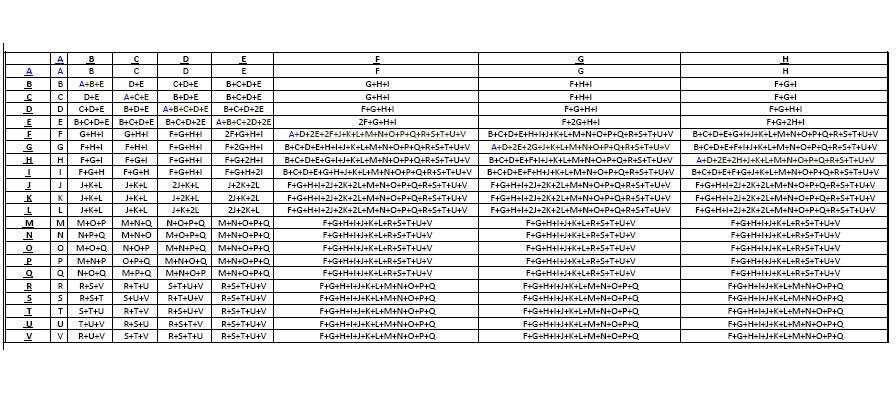}}
\caption[$A_5$]{Fusion Rules for $A_5$}
\label{fig:A5a}
\end{figure}

\begin{figure}[t]
\centerline{\includegraphics[width=19cm, height=9cm]{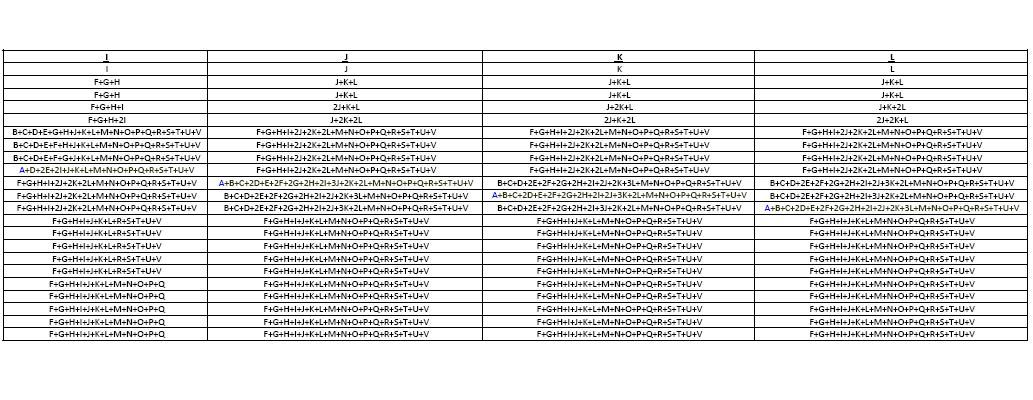}}
\caption[$A_5$]{Fusion Rules for $A_5$}
\label{fig:A5b}
\end{figure}

\begin{figure}[t]
\centerline{\includegraphics[width=19cm, height=9cm]{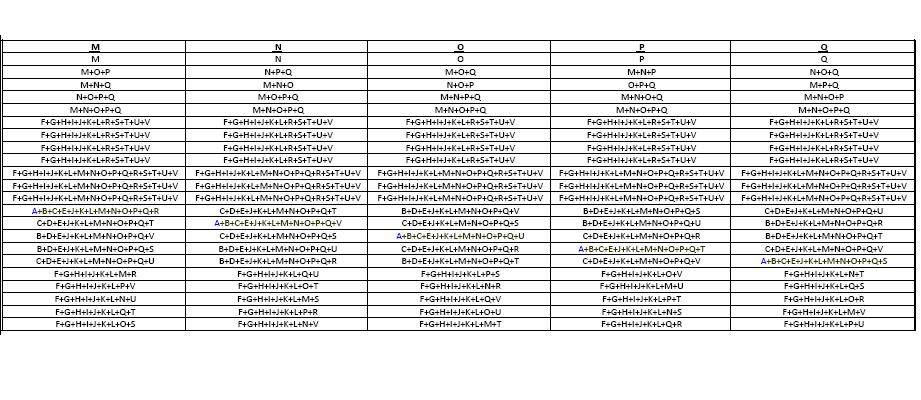}}
\caption[$A_5$]{Fusion Rules for $A_5$}
\label{fig:A5c}
\end{figure}

\begin{figure}[t]
\centerline{\includegraphics[width=19cm, height=9cm]{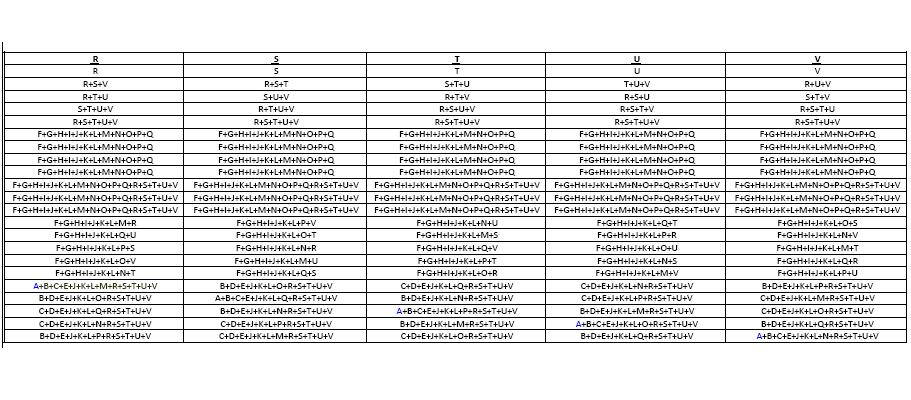}}
\caption[$A_5$]{Fusion Rules for $A_5$}
\label{fig:A5d}
\end{figure}

\section{MAGMA Program and Discussion of Results}

The tables computed in this paper were produced using the MAGMA program included in Appendix 1.  The MAGMA program
computes the particles (and their properties - e.g. spin), braiding, and fusion rules for an arbitrary finite residual gauge group.  The program can also explicitly compute the entries of the S matrix.  As part of the fusionprinter(Group) procedure, the MAGMA program computes the change of basis S matrix and implements the Verlinde formula to determine the fusion coefficients based on the S matrix for an arbitrary group.

In addition, the MAGMA program can compute the probabilities of obtaining any given fusion product used in many different topological quantum computation schemes.  More detail on this connection is provided in the next section.

The two examples included here were chosen mainly due to their possible applications to universal quantum
computation (as described in the next section).  In particular, since $S_3$ is the smallest non-abelian group, a method for achieving universal quantum computation in that case has the best chance of being engineered in the lab or found in nature.  
In Appendix B, the quantum doubles for the remaining finite subgroups of $SO(3)$ are tabulated and analyzed.  Each of these are interesting from either a physics theory or implementation point of view.  
For example, from the physics theory
perspective, in the two theories $S_3$ and $A_5$ discussed above, every particle is its
own antiparticle.  However, this is not always the case and $A_4$ 
illustrates this point.
Furthermore, other dihedral groups, for example, could be interesting from the implementation
perspective as they arise naturally from spontaneous symmetry breaking in p-wave superconductors.

\subsection{Majorana Particles}

From the fusion tables derived for $S_3$ and $A_5$, every particle is its own antiparticle, i.e. the particles are all Majorana.  Most commonly, photons are their own antiparticles.  However, Majorana particles of noninteger spin are currently an active area of research in physics, even beyond condensed matter applications.  In particular, neutrinos are conjectured by some to be Majorana and SUSY would demand a fermionic photino as the fermionic superpartner of the photon (among many other Majorana particles).  Finally, one suggestion is that WIMPS (a dark matter candidate, possibly the dominant component of dark matter mass in our universe) are also Majorana.

\section{Quantum Computation Schemes}

\subsection{Universal Quantum Computation}

Universal computation: any unitary operation can be approximated to arbitrary accuracy

\subsection{Schemes for Topological Quantum Computation}

States used in a topological quantum computation scheme must be selected according to their fusion rules since the environment can distinguish between states fusing to different superselection sectors, causing decoherence.  
In addition, most implementation proposals use particle-antiparticle pairs as the fundamental building block since these interact trivially with other pairs when kept far apart from each other.  Braiding computations are achieved by transporting a given pair in between other pair(s).  

If anyons from different superselection sectors are used in the actual quantum computation scheme, then two distance scales come into play.  In particular, the anyons must be kept far enough apart from each other that they cannot influence each other by local interactions (and thus superselection sectors are exhibited).  Similarly, the environment must be far enough removed that it can only detect the cumulative long range quantum numbers of the whole system. (This provides protection against decoherence.)

The prototype of anyon computation based on superselection sectors is the Fibonacci (also known as Yang-Lee) anyons, in which there are only two superselection sectors 0 and 1, with nontrivial fusion rule $1\times1=0+1$.  The dimension of the Hilbert space fusing to the vacuum increases exponentially in the golden ratio $\phi=\dfrac{1+\sqrt{5}}{2}$.  The associated R-matrix and its F-matrix conjugate $FR F^{-1} $ then form a dense subgroup of $SU(2)$.  Thus any unitary operation can be approximated with arbitrary accuracy and universal quantum computation is possible.  The Fibonacci anyons have spin given by $\frac{4}{5}$ (This can be obtained by, for instance, solving the pentagon and hexagons explicitly for the R matrix and then using the generalized spin-statistics connection to infer the spins.)  Thus this model is equivalent to $SO(3)_3$ (the restriction of $SU(2)_3$).

\subsubsection{Fusion Probabilities}

Finally, the probability results obtained for Fibonacci anyons in~\cite{PreskillCh9} can be extended to anyons based on the quantum groups of finite groups as well.  Adopting Preskill's diagram convention of weighting by a compensating factor of $\sqrt{d_A} $ for each anyon-antianyon kink in a particle A's worldline, the dimension becomes equal to that of a topologically equivalent worldline.  (Ordinarily every anyon-antianyon worldline zigzag would generate a factor of $\dfrac{1}{d_A}$ where $d_A$ denotes the quantum dimension of particle A.)

Consider the topological quantum computation schemes in which particle-antiparticle pairs are created from the vacuum.  (This is standard in Kitaev's, Preskill's and Mochon's schemes.)  Using the above diagram convention, Figures [9] and [10] 
show that the following formula for the probability that a particular fusion product occurs holds:

$$ \mathbb{P}(A B \rightarrow C) = \dfrac{N_{AB}^C d_C}{d_A d_B} $$

\begin{figure}[t]
\centerline{\includegraphics[width=11cm, height=6cm]{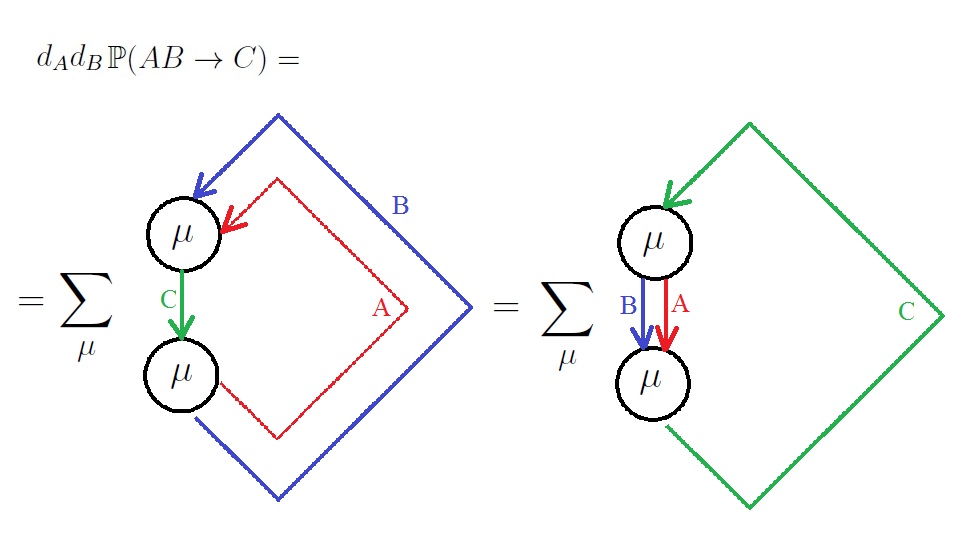}}
\caption[Fusion Probabilities]{Fusion Probabilities.} %
\label{fig:probderiv}
\end{figure}

\begin{figure}[t]
\centerline{\includegraphics[width=8cm, height=4cm]{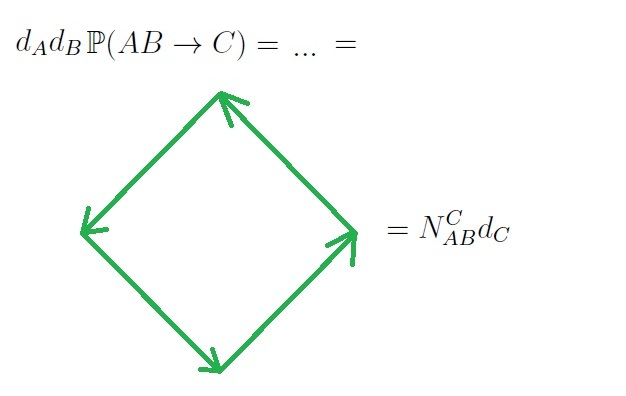}}
\caption[Fusion Probabilities]{Fusion Probabilities Cont'd.} %
\label{fig:probderivb}
\end{figure}

\subsection{Possible Realizations of Topological Quantum Computation}

\subsubsection{$A_5$}
Ogburn and Preskill showed that $A_5=PSL(2, \F_4)$ is sufficient for universal quantum computation under the assumption that
a supply of magnetic ancillas is available (but not necessarily electric ancillas).~\cite{ogburnpreskill}  This sufficiency condition was subsequently generalized to any nonsolvable finite group.~\cite{unimoch}  Note that geometrically $A_5$ is the subgroup of rotations of the icosahedral group (the symmetries of a regular icosahedron - one of the five platonic solids), which in turn is a subgroup of $SO(3)$.  Out of the class of nonsolvable finite groups, $A_5$ has the best chance of being engineered in the lab since it has the smallest order (60 elements).  It is unlikely to find a symmetry group with such high order in nature in a useful form.

\subsubsection{$S_3$}

Since the smallest nonsolvable group is $A_5$ and $|A_5|=60$, it is desirable to analyze smaller groups.  $S_3=D_3$ is the smallest nonabelian group ($|S_3| = 6 = \frac{|A_5|}{10}$), so in that sense a quantum computation scheme based on $S_3$ would be optimal.  $S_3$ symmetry groups arise naturally in two dimensional lattices as the symmetry group of a hexagonal lattice (for example, graphene) so it is perhaps more reasonable to expect to find an $S_3$ system conducive to topological gauge theories in nature.  In addition, proposals have been suggested for engineering $S_3$ in optical lattices or ion traps.

Since there are 3-particle nonabelian subsystems of $S_3$, one might hope that one would lead to universal quantum computation in a manner similar to the Fibonacci anyons.  Unfortunately, the subsystems do not generate an SU(2)-dense 3 strand representation of the braid group as the $0$ and $1$ particles do in the case of the Fibonacci anyons.  In particular, consider a 3-particle subsystem of $S_3$.  Note that if we could also get rid of the B particle, then we obtain a subsystem with fusion rules given by the Fibonacci anyons.  Unfortunately, the spins, and thus the R matrices, are not the same as in the Fibonacci case.  In general, the $SU(2)_k$ particles have spins that are specific multiples of $\frac{1}{k+2}$ (phase factors of $e^{\frac{2 \pi i}{k+2}} $).

For the moment, let P denote the particle with the nontrivial fusion rule (in all cases, the dimension of the internal Hilbert space of P is 2).  $N^A_{nP}$, where $nP$ represents $n$ $P$ particles, satisfies the recursion $ N^A_{n,P} =2*N^A_{n-2,P}+N^A_{n-1,P} $.  The solution to this recursion is $N^A_{n,P}= \dfrac{(1+\sqrt{2})^n - (1-\sqrt{2})^n}{2 \sqrt{2}} $.  Thus the resulting subsystem has an exponentially increasing Hilbert space dimension.  In the case of $P=G$ or $P=H$, the particles have spin $\frac{1}{3}$ or $\frac{2}{3}$ and the fusion rules correspond to those of $SO(3)_4$ or its parity image.  Unfortunately, $SO(3)_4$ is the restriction of $SU(2)_4$ and it has been shown that $SU(2)_4$ does not yield universal quantum computation.~\cite{FreedTEVP}  $P=C$ or $P=F$ have spin $0$, so do not even have nontrivial R matrices.  Thus the braid group representation described for the Fibonacci anyons would be trivial when considering combinations of $F$ anyons.

A method for universal quantum computation based on qutrits (as opposed to qubits) in $S_3$ that uses more than 3 particle types (including D fluxes and electric charges in addition to the vacuum A) was provided by Kitaev.  The construction was generalized to solvable but non-nilpotent groups.~\cite{smallermoch}  The major drawback of the method relative to the $A_5$ case is that it requires creation and fusion of electric charges, which may be more challenging than magnetic fluxes to create.  In particular, the computational power of just conjugation alone will not give a Toffoli gate in the case of $S_3$ as it does in the case of $A_5$.

\subsubsection{The Dihedral Groups and $ p + i p$ Superconductors}

From the field theory point of view, time reversal or parity symmetry is broken in chiral p-wave superconductors so the electromagnetic $U(1)$ group becomes minimally non-Abelian - $U(1) \ltimes \Z/2\Z$.  Superconductivity results when this gauge group is spontaneously broken, and could give rise to non-Abelian anyons as, for instance, $D_n = \Z/n\Z \ltimes \Z/2\Z$.


\section{Conclusions}

In this paper, (2+1)-dimensional topological gauge theories based on finite gauge theories are analyzed with a view towards physical implementations and quantum computation.  A program to compute the particles, their properties, the S matrix, and the fusion rules for a general finite group was written.  The fusion rules for the quantum doubles most physically or computationally relevant were tabulated and the resulting tables were analyzed.  In particular, the particles, fusion rules, and subsystems for two groups thought to be sufficient for universal quantum computation under certain circumstances - $A_5$ and $S_3$ - were analyzed in some detail.  The $S_3$ theory with 8 particles was found to have many three particle subsystems.  Unfortunately, these either led to trivial R matrices or a system given by $SO(3)_4$, which has been shown to not lead to universal quantum computation since it does not produce a dense braid group representation.  In addition, the particles, properties, and fusion rules for $A_5$ were tabulated and analyzed.  The tabulation of the quantum doubles of the next dihedral subgroup and all non-dihedral finite subgroups of $SO(3)$ is completed in the appendices.

\begin{acknowledgments}
 The author would like to thank F. Wilczek for advising her thesis (the flagstone for this paper) and for subsequently suggesting she ``condense a paper out of it".  In addition, she would like to thank P. Shor for providing access to MAGMA and helping to debug part of the MAGMA code.

\end{acknowledgments}

\bibliographystyle{ieeetr}
\bibliography{arxiv}

\begin{thebibliography}{10}

\bibitem{Feynman}
R.~Feynman, ``Simulating physics with computers,'' {\em {International Journal
  of Theor.}}, vol.~{Physics 21}, p.~467, 1982.

\bibitem{shor}
P.~Shor, ``Polynomial-time algorithms for prime factorization and discrete
  logarithms on a quantum computers,'' {\em {SIAM J. Comput.}}, vol.~{\bf
  26(5):}, pp.~1484--1509, 1997.
\newblock preprint {\tt arXiv:quant-ph/9508027v2}.

\bibitem{steane}
A.~Steane, ``Multiple-particle interference and quantum error correction,''
  {\em {Proceedings of the Royal Society A}}.

\bibitem{Or}
D.~Aharonov and M.~Ben-Or, ``Fault-tolerant quantum computation with constant
  error rate,'' June 1999.
\newblock preprint {\tt arXiv:quant-ph/9906129v1}.

\bibitem{ady}
A.~Stern, ``Anyons and the quantum hall effect - a pedagogical review,'' Nov.
  1997.
\newblock preprint {\tt arXiv:0711.4697v1}.

\bibitem{wil}
F.~Wilczek, ``Quantum mechanics of fractional-spin particles,'' {\em {Phys.
  Rev. Lett.}}, vol.~{\bf 49}, p.~14, 1982.
\newblock preprint {\tt arXiv:quant-ph/}.

\bibitem{weeks}
J.~Hoste, M.~Thisthlethwaite, and J.~Weeks, ``The first 1,701,936 knots,'' {\em
  {The Mathematical Intelligencer}}, vol.~{\bf 20}, p.~33.

\bibitem{prop}
M.~de~Wild~Propitius and F.~A. Bais, ``Discrete gauge theories,'' Mar. 1996.
\newblock preprint {\tt arxiv hep-th/9511201 v2}.

\bibitem{aharonovbohm}
Y.~Aharonov and D.~Bohm, ``Significance of electromagnetic potentials in the
  quantum theory,'' {\em {The Physical Review}}, vol.~{\bf 115}, p.~485.

\bibitem{metalring}
A.~van Oudenaarden, M.~H. Devoret, Y.~V. Nazarov, and J.~E. Mooij,
  ``Magnetoelectric aharonov-bohm effect in metal rings,'' {\em {Nature}},
  vol.~{\bf 391}, p.~768.

\bibitem{wein}
{Steven Weinberg}, ``The quantum theory of fields,'' 1995.

\bibitem{cole}
{Sidney Coleman}, ``Aspects of symmetry: Selected erice lectures,'' 1985.

\bibitem{higgscreen}
F.~A. Bais, A.~Morozov, and M.~de~Wild~Propitius, ``Charge screening in the
  higgs phase of chern-simons electrodynamics,'' {\em {Physical Review
  Letters}}, vol.~{\bf 71}, p.~2383.

\bibitem{haag}
K.~Druhl, R.~Haag, and J.~Roberts, ``On parastatistics,'' {\em {Commun. math.
  Phys.}}, vol.~{\bf 18}, p.~204.

\bibitem{koorn}
T.~Koornwinder, B.~Schroers, J.~Slingerland, and F.~Bais, ``Fourier transform
  and the verlinde formula for the quantum double of a finite group,'' {\em
  {jpa}}, vol.~{\bf 32}, p.~8539.

\bibitem{kitaevToric}
A.~Kitaev, ``Fault-tolerant quantum computation by anyons,'' {\em {Annals Phys.
  303, 2-30}}, May 2003.
\newblock preprint {\tt arXiv:quant-ph/9707021v1}.

\bibitem{luk}
L.~Duan, E.~Demler, and M.~Lukin, ``Controlling spin exchange interactions of
  ultracold atoms in optical lattices,'' {\em {Physical Review Letters}},
  vol.~{\bf 91}, p.~9, Aug. 2003.
\newblock preprint {\tt arXiv:cond-mat/0210564v2}.

\bibitem{JJ}
B.~Doucot, L.~Ioffe, and J.~Vidal, ``Discrete non-abelian gauge theories in
  two-dimensional lattices and their realizations in josephson-junction
  arrays,'' {\em {Physical Review B}}, vol.~{\bf 69}, 2004.

\bibitem{PreskillCh9}
J.~Preskill, ``Chapter 9: Topological quantum computation,''
\newblock online notes.

\bibitem{ogburnpreskill}
R.~W. Ogburn and J.~Preskill, ``Topological quantum computation,'' in {\em
  QCQC98} (C.~Williams, ed.), (Berlin), pp.~341--356, Springer Verlag, 1999.

\bibitem{unimoch}
C.~Mochon, ``Anyons from nonsolvable finite groups are sufficient for universal
  quantum computation,'' {\em {Phys. Rev. A}}, vol.~{\bf 67}, p.~022315, 2003.

\bibitem{FreedTEVP}
M.~Freedman, M.~Larsen, and Z.~Wang, ``The two-eigenvalue problem and density
  of jones representation of braid groups,'' {\em {Commun. Math. Phys. 228,
  177-199}}, 2003.

\bibitem{smallermoch}
C.~Mochon, ``Anyon computers from smaller finite groups,''
\newblock preprint {\tt arXiv:quant-ph/0306063 v2}.

\end{thebibliography}

\appendix

\section{Appendix}

{\tt
\begin{verbatim}
function ComputeSMatrix(A, B, alpha, beta, Gp)
CC:=ConjugacyClasses(Gp);
Cent := [Centralizer(Gp,Gp!CC[i][3]) : i in [1..#CC]];
total:=0;
for j in [1..#RightTransversal(Gp,Cent[B])] do
for i in [1..#RightTransversal(Gp,Cent[A])] do
cj:= RightTransversal(Gp, Cent[B])[j];
ci:= RightTransversal(Gp, Cent[A])[i];
cj:=cj^-1;
ci:=ci^-1;
valpha := ci^-1 * cj * CC[B][3] * cj^-1 * ci;
valbeta := cj^-1 * ci * CC[A][3] * ci^-1 * cj;
if (Gp!cj * CC[B][3] * cj^-1)^-1 * (Gp! ci * CC[A][3] * ci^-1)^-1
    *(Gp!cj * CC[B][3] * cj^-1) * (Gp! ci * CC[A][3] * ci^-1) eq Id(Gp)
then total:=total+ComplexConjugate((CharacterTable(Cent[A])[alpha])(valpha))*
    ComplexConjugate((CharacterTable(Cent[B])[beta])(valbeta));
end if;
end for;
end for;
total:=total/#(Gp);
return(total);
end function;

function FusionCoefficientsviaSMatrix(A, alpha, B, beta, C, gamma, Gp)
CC:=ConjugacyClasses(Gp);
Cent := [Centralizer(Gp,Gp!CC[i][3]) : i in [1..#CC]];
fusionsum:=0;
//sum over all centralizers D and representations of those centralizers delta
for D in [1..#ConjugacyClasses(Gp)], delta in [1..#ConjugacyClasses(Cent[D])] do
//since the number of conjugacy classes = number of representations
//want delta up to # of repns of the centralizer corresponding to repn D Cent[D]
if(ComputeSMatrix(1,D,1,delta,Gp) ne 0)
then
fusionsum:=fusionsum+ComputeSMatrix(A, D, alpha, delta, Gp) *
ComputeSMatrix(B, D, beta, delta, Gp) *
ComplexConjugate(ComputeSMatrix(C, D, gamma, delta, Gp))/
    ComputeSMatrix(1, D, 1, delta, Gp);
end if;
end for;
return fusionsum;
end function;

function FusionProb(A, alpha, B, beta, C, gamma, Gp)
CC:=ConjugacyClasses(Gp);
DA:=CharacterTable(Centralizer(Gp, Gp!CC[A][3]))[alpha][1];
dimA:= (CC[A][2])*DA;
DB:=CharacterTable(Centralizer(Gp, Gp!CC[B][3]))[beta][1];
dimB:= (CC[B][2])*DB;
DC:=CharacterTable(Centralizer(Gp, Gp!CC[C][3]))[gamma][1];
dimC:= (CC[C][2])*DC;
fus:=FusionCoefficientsviaSMatrix(A, alpha, B, beta, C, gamma, Gp);
prob:= ((fus) * (dimC))/(dimA *dimB);
return(prob);
end function;

function fusionprinter(Gp)
CC:=ConjugacyClasses(Gp);
Cent := [Centralizer(Gp,Gp!CC[i][3]) : i in [1..#CC]];

for i in [1..#ConjugacyClasses(Gp)],
i1 in [1..#ConjugacyClasses(Cent[i])]  do
printf " \n";
for j in [1..#ConjugacyClasses(Gp)],
j1 in [1..#ConjugacyClasses(Cent[j])] do
printf ", ";
firstprint := 1;
particlenum := 0;
for k in [1..#ConjugacyClasses(Gp)], k1 in
[1..#ConjugacyClasses(Cent[k])] do
particlenum := particlenum + 1;
if (FusionCoefficientsviaSMatrix(i, i1, j, j1, k, k1,Gp) ne 0) then
   if firstprint eq 0 then printf "+";
   end if;
   firstprint := 0;
   if (FusionCoefficientsviaSMatrix(i, i1, j, j1, k, k1,Gp) ge 2) then
   printf "\%o.", FusionCoefficientsviaSMatrix(i, i1, j, j1, k, k1,Gp);
   end if;
   printf "\%o", particlenum;
end if;
end for;
end for;
end for;

return 0;
end function;

function spinprinter(Gp, A, alpha)  //returns spin mod 1
CC:=ConjugacyClasses(Gp);
CT:=CharacterTable(Centralizer(Gp, Gp!CC[A][3]))[alpha];
C := ComplexField();
z:=(C!CT(CC[A][3]))/(C!CT[1]);
s1:=Log(z)/(2*Pi(C)*C.1);
s:=(s1+ComplexConjugate(s1))/2;
return(s);
end function;

\end{verbatim}
}

%
%
%
%
%
\section{Other Interesting Quantum Doubles}
\label{ap:other}

\subsection{Finite Subgroups of $SO(3)$ and $SU(2)$}

 In this appendix, the tabulation of the quantum doubles of all non-dihedral finite subgroups of $SO(3)$ is completed.  
In the case of the infinite family $D_n$, the quantum doubles for the smallest order groups are tabulated. (These are most likely to be physically relevant.)

In addition to being potentially physically relevant, the tables for the remaining subgroups of $SO(3)$ provide a demonstration of the variety of properties that can be exhibited by a quantum double.  Specifically, in the $A_4$ quantum double, not every particle is Majorana as it is in $S_3$ and $A_5$.




%
%
%
%


\subsection{$A_4$}

$A_4=T$, the symmetries of the tetrahedron, is the next smallest alternating group.  Representatives for the conjugacy classes are given by: Id, (12)(34) (3 elements in the conjugacy class), (123) (4 elements in the conjugacy class), and (132) (4 elements in the conjugacy class) - Their centralizers are $A_4 = <(12)(34), (234)>$, $\Z/2\Z \times \Z/2\Z=<(13)(24),(12)(34)>$, $\Z/3\Z = <(123)>$, and $\Z/3\Z=<(123)>$.  The character tables of the centralizers were all given in the text for another group, except of course $A_4$ itself. The character table for $A_4$ is:

\begin{center}
\begin{tabular}[t]{l|c|c|c|c}
\multicolumn{5}{c}{Representations of $A_4$}\\\hline
Repn & (1) & (12)(34) & (123) & (132) \\ \hline
$\kappa_1$  & 1 & 1 & 1 & 1 \\
$\kappa_2$  & 1 & 1&-1-$\zeta$ &  $\zeta$ \\
$\kappa_3$  & 1 & 1 &  $\zeta$ & -1-$\zeta$ \\
$\kappa_4$  & 3 & -1 & 0 & 0 \\

\end{tabular}
\end{center}

where $\zeta^2+\zeta+1=0$.

Particles are specified by flux (a choice of conjugacy class) and charge (specified by an irreducible representation of the centralizer of a representative element of the flux conjugacy class). Thus there are 14 particles, given by:

\begin{center}
  \begin{tabular}{ c | c | c | c | c | c}
    Particle & Flux & Charge & Spin & Repn Dim & Dim Internal Hilbert Space \\    \hline
   A & () & $\kappa_1$ & 0  & 1 & 1  \\
   B & () & $\kappa_2$ & 0 & 1 & 1\\
   C & () & $\kappa_3$ & 0 & 1 & 1\\ 
   D & () & $\kappa_4$ & 0 & 3 & 3\\ 
   E & (12)(34) & $\eta_1$ & 0 & 1 & 3 \\
   F & (12)(34) & $\eta_2$ & 0 & 1 & 3 \\
   G & (12)(34) & $\eta_3$ & $\frac{1}{2}$ & 1 & 3\\
   H & (12)(34) & $\eta_4$ & $\frac{1}{2}$ & 1 & 3\\
   I & (123) & $\tilde\rho_1$ & 0 & 1 & 4\\
   J & (123) & $\tilde\rho_2$ & $\frac{1}{3}$ & 1 & 4 \\
   K & (123) & $\tilde\rho_3$ & $\frac{2}{3}$ & 1 & 4 \\
   L & (132) & $\tilde\rho_1$ & 0 & 1 & 4 \\
   M & (132) & $\tilde\rho_2$ & $\frac{2}{3}$ & 1 & 4\\
   N & (132) & $\tilde\rho_3$ & $\frac{1}{3}$ & 1 & 4\\
  \end{tabular}
\end{center}

Note that the sum of the squares of the internal Hilbert spaces is $3*1^2+3^2+4*3^2+6*4^2=144=|A_4|^2$, the dimension of the quantum double.

\begin{figure}[t]
\centerline{\includegraphics[width=8 in, height=5 in]{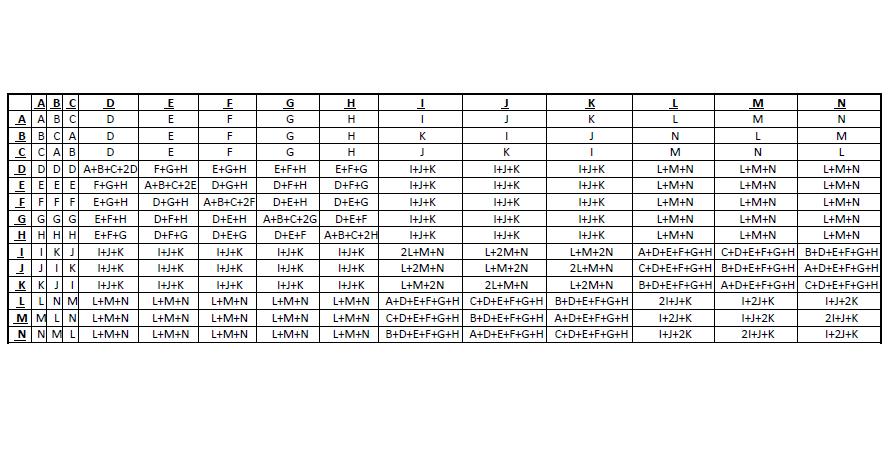}}
\caption[$A_4$]{Fusion Rules for $A_4$}
\label{fig:A4}
\end{figure}

Note that B and C are antiparticles, so this is the first example we have of a theory in which not every particle is its own antiparticle.

\subsection{$S_4$}%

The next highest order symmetric group is $S_4$, with $4!=120$ elements.  There are five conjugacy classes (specified by the cycle decomposition) - Id, (12), (12)(34), (123), and (1234).  The centralizers for these conjugacy classes are, respectively, $S_4$, $D_4$, $\Z/ 2\Z \times \Z/2\Z $, $\Z/3\Z$, and $\Z/4\Z$.  The character tables for all of these groups (with the exception of $S_4$ and $D_4 $) have been given in the main text, so will not be repeated here (the same representation labels will be used in the enumeration of the particles).

The character table for $S_4$ is:

\begin{center}
  \begin{tabular}{ c | c | c | c | c | c}
Irrep & () & (13)(24) & (12) & (123) & (1234) \\    \hline
$\tilde\chi_1$   &   1 & 1&  1 & 1 & 1 \\
$\tilde\chi_2$   &   1 & 1& -1 & 1 &-1 \\
$\tilde\chi_3$   &   2 & 2&  0 &-1 & 0 \\
$\tilde\chi_4$   &   3 &-1& -1 & 0 & 1 \\
$\tilde\chi_5$   &   3 &-1&  1 & 0 & -1 \\
  \end{tabular}
\end{center}

The character table of $D_4$ is:

\begin{center}
  \begin{tabular}{ c | c | c | c | c | c}
Irrep & () & (13)(24) & (14)(23) & (24) & (1234) \\    \hline
$\Upsilon_1$ & 1 & 1 & 1 & 1 & 1 \\
$\Upsilon_2$ & 1 & 1 & 1 & -1 & -1 \\
$\Upsilon_3$ & 1 & 1 & -1& -1 & 1 \\
$\Upsilon_4$ & 1 & 1 & -1 & 1 & -1 \\
$\Upsilon_5$ & 2 & -2 & 0 & 0 & 0 \\
  \end{tabular}
\end{center}

There are 21 particles.


\begin{center}
  \begin{tabular}{ c | c | c | c}  
    Particle & Flux & Charge & Spin   \\ \hline 

A & () & $\tilde\chi_1$ & 0 \\
B & () & $\tilde\chi_2$ & 0 \\
C & () & $\tilde\chi_3$ & 0 \\
D & () & $\tilde\chi_4$ & 0 \\
E & () & $\tilde\chi_5$ & 0 \\
F & (12)(34) & $\Upsilon_1$ & 0 \\
G & (12)(34) & $\Upsilon_2$ & 0 \\
H & (12)(34) & $\Upsilon_3$ & 0 \\
I & (12)(34) & $\Upsilon_4$ & 0 \\
J & (12)(34) & $\Upsilon_5$ & $\frac{1}{2}$ \\
K & (12) & $\eta_1$ & 0 \\
L & (12) & $\eta_2$ & 0 \\
M & (12) & $\eta_3$ & $\frac{1}{2}$ \\
N & (12) & $\eta_4$ & $\frac{1}{2}$ \\
O & (123) & $\tilde\rho_1$ & 0 \\
P & (123) & $\tilde\rho_2$ & $\frac{1}{3}$ \\
Q & (123) & $\tilde\rho_3$ & $\frac{2}{3}$ \\
R & (1234) & $\tilde{\tilde\rho}_1$ & 0 \\
S & (1234) & $\tilde{\tilde\rho}_2$ & $\frac{1}{4}$ \\
T & (1234) & $\tilde{\tilde\rho}_3$ & $\frac{1}{2}$ \\
U & (1234) & $\tilde{\tilde\rho}_4$ & $\frac{3}{4}$ \\

\end{tabular}
\end{center}

\begin{figure}[t]
\centerline{\includegraphics[width=6 in, height=3.5 in]{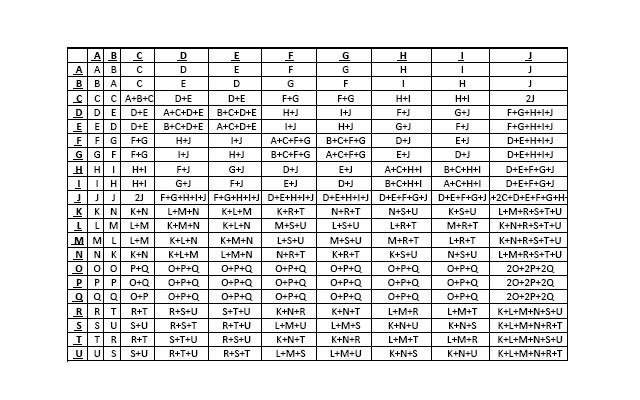}}
\caption[$S_4$]{Fusion Rules for $S_4$}
\label{fig:S4a}
\end{figure}

\begin{figure}[t]
\centerline{\includegraphics[width=6 in, height=3.5 in]{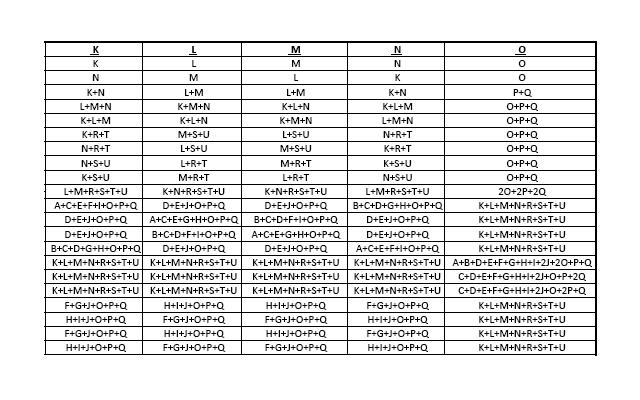}}
\caption[$S_4$]{Fusion Rules for $S_4$}
\label{fig:S4b}
\end{figure}

\begin{figure}[t]
\centerline{\includegraphics[width=6 in, height=3.5 in]{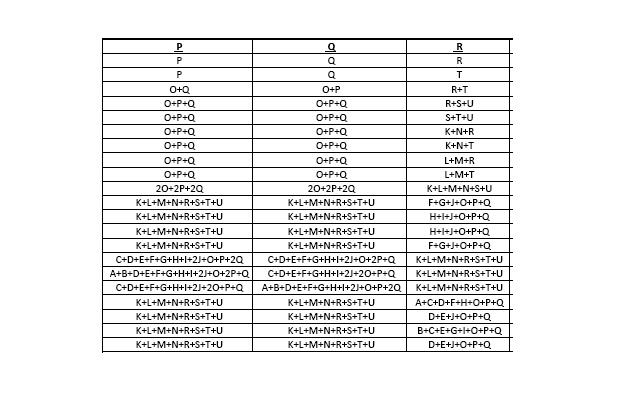}}
\caption[$S_4$]{Fusion Rules for $S_4$}
\label{fig:S4c}
\end{figure}

\begin{figure}[t]
\centerline{\includegraphics[width=6 in, height=3.5 in]{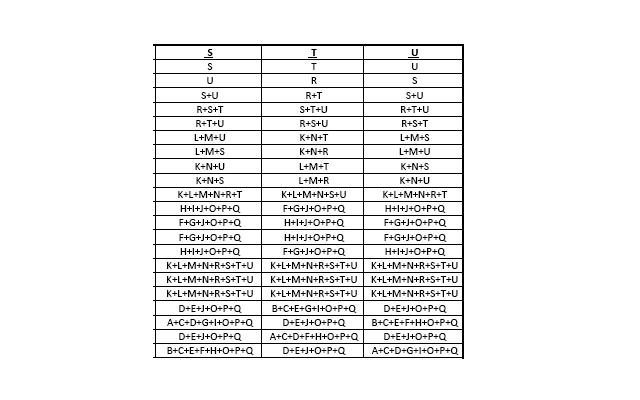}}
\caption[$S_4$]{Fusion Rules for $S_4$}
\label{fig:S4d}
\end{figure}

From the fusion rule table, every particle is its own antiparticle.

\subsection{$D_4$}

The conjugacy classes are given by Id, (13)(24) (1 element in the conjugacy class), (14)(23) (2 elements in the conjugacy class), (24) (2 elements in the conjugacy class), (1234) (2 elements in the conjugacy class) - The centralizers are respectively $D_4$, $D_4$, $\Z/2\Z \times \Z/2\Z = <(12)(34),(24)>$, $\Z/2\Z \times \Z/2\Z = <(13),(24)>$, $\Z/4\Z = <(1234)>$.

Thus there are 22 particles.

\begin{center}
  \begin{tabular}{ c | c | c | c | c}
    Particle & Flux & Charge & Repn Dim & Dim Internal Hilbert Space \\    \hline
   A & () & $\Upsilon_1$ & 1 & 1  \\
   B & () & $\Upsilon_2$ & 1 & 1\\
   C & () & $\Upsilon_3$ & 1 & 1\\ 
   D & () & $\Upsilon_4$ & 1 & 1\\ 
   E & () & $\Upsilon_5$ & 2 & 2 \\
   F & (13)(24) & $\Upsilon_1$ & 1 & 1 \\
   G & (13)(24) & $\Upsilon_2$ & 1 & 1\\
   H & (13)(24) & $\Upsilon_3$ & 1 & 1\\
   I & (13)(24) & $\Upsilon_4$ & 1 & 1\\
   J & (13)(24) & $\Upsilon_5$ & 2 & 2 \\
   K & (14)(23) & $\eta_1$ & 1 & 2 \\
   L & (14)(23) & $\eta_2$ & 1 & 2 \\
   M & (14)(23) & $\eta_3$ & 1 & 2\\
   N & (14)(23) & $\eta_4$ & 1 & 2\\
   O & (24) & $\eta_1$ & 1 & 2\\
   P & (24) & $\eta_2$ & 1 & 2\\
   Q & (24) & $\eta_3$ & 1 & 2\\
   R & (24) & $\eta_4$ & 1 & 2\\
   S & (1234) & $\tilde{\tilde\rho}_1$ & 1 & 2\\
   T & (1234) & $\tilde{\tilde\rho}_2$ & 1 & 2\\
   U & (1234) & $\tilde{\tilde\rho}_3$ & 1 & 2\\
   V & (1234) & $\tilde{\tilde\rho}_4$ & 1 & 2\\
  \end{tabular}
\end{center}

As a check, note that the sum of the dimensions of the internal Hilbert spaces of the particles is $8*1^2+14*2^2=64=|D_4|^2$, the dimension of the quantum double.

\begin{figure}[t]
\centerline{\includegraphics[width=6 in, height=3.25 in]{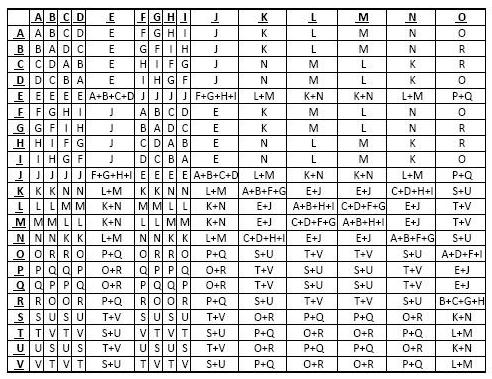}}
\caption[$D_4$]{Fusion Rules for $D_4$}
\label{fig:D4a}
\end{figure}

\begin{figure}[t]
\centerline{\includegraphics[width=4.5 in, height=3.25 in]{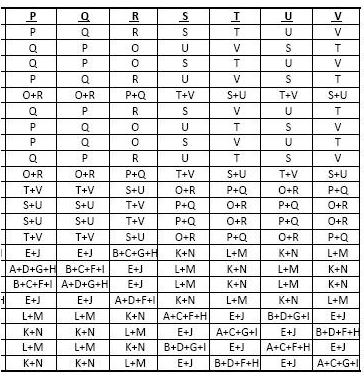}}
\caption[$D_4$]{Fusion Rules for $D_4$}
\label{fig:D4b}
\end{figure}

Note by inspection that each particle is again its own antiparticle in this $D_4$ theory.

\subsubsection{Possible Physical Realization}

Note that $D_4 = \Z/4\Z \ltimes \Z/2\Z$, so one possible physical realization could arise from SSB in chiral p-wave superconductors.

\end{document}